\documentclass[12pt]{article}
\usepackage{amsmath,amssymb,amsfonts,color,graphicx,cite,color,soul}
\usepackage{epsfig}
\usepackage{blindtext}
\usepackage[utf8]{inputenc}
\usepackage[T1]{fontenc}
\usepackage{caption}
\usepackage{subcaption}
\usepackage{capt-of}
\usepackage{hyperref}
\usepackage{latexsym,amstext,amscd}
\usepackage{setspace}
\usepackage{enumerate}
\usepackage{epic}
\usepackage{mathrsfs}
\usepackage{breqn}
\usepackage{float}
\usepackage{multirow}
\usepackage{bbm}
\usepackage{verbatim}
\usepackage[vcentering,dvips]{geometry}
\usepackage{blindtext}
\usepackage{abstract,lipsum}
\usepackage{breqn}

\usepackage[bottom]{footmisc}
\usepackage[affil-it]{authblk}
\usepackage[usenames]{xcolor}
\usepackage[english]{babel}
\usepackage{dcolumn}

\newcommand{\nn}{\nonumber}

\input paperdef

\title{Flavour in $SU(5)$ Finite Grand Unified Models}
\author{Luis Od\'in Estrada Ramos$^{1}$\thanks{email: o.estramos@ciencias.unam.mx}~,
Myriam Mondrag\'on$^{1}$\thanks{email: myriam@fisica.unam.mx}~,
Gregory Patellis$^{2}$\thanks{email: grigorios.patellis@tecnico.ulisboa.pt}
and George Zoupanos$^{3,4}$\thanks{email: george.zoupanos@cern.ch}\\
{\small
$^1$Instituto de F\'{\i}sica, Universidad Nacional Aut\'onoma de M\'exico, A.P. 20-364, CDMX 01000 M\'exico\\ %
$^2$ Centro de Física Teórica de Partículas - CFTP, Departamento de Física,\\ Instituto Superior Técnico, Universidade de Lisboa,\\
Avenida Rovisco Pais 1, 1049-001 Lisboa, Portugal \\
$^3$ Physics Department,   National Technical University, 157 80 Zografou, Athens, Greece\\
$^4$ Max-Planck Institut f\"ur Physik, F\"ohringer Ring 6, D-80805
  M\"unchen, Germany 
}
}\date{}

\begin{document}

\maketitle

\begin{abstract}
{Four $SU(5)$ $N=1$ supersymmetric models which exhibit $S_3$ and/or $Z_N$ symmetries are studied, that are finite to two or all loops, and their corresponding mass matrices. The first is an all-loop finite model based on an $S_3\times Z_3\times Z_2$ flavour symmetry, which leads to phenomenologically nonviable mass matrices. The remaining models, based  on cyclic symmetries, show various mass textures, some of which are phenomenologically promising.  
For the two-loop finite models, the parametric solutions to the finiteness conditions determine completely some of the Yukawa couplings, and lead to a restricted range of values for other ones at the GUT scale, with a considerable reduction in the number of free parameters. One particular solution of the two-loop models shows an enhanced symmetry, leading to an all-loop finite model, which has a significant parameter reduction and could in principle reproduce the observed quark masses and mixing pattern. In this case the finiteness conditions determine the absolute value of all the Yukawa couplings at the unification scale. Finally, the minimum number of phases in the mass matrices and their position are determined, a task not previously done in Finite Unified Theories, which contributes towards the reduction of parameters and a better understanding of the Yukawa couplings.} 
\end{abstract}
\maketitle

\section{Introduction}
 
A fundamental theory of nature should be able to give predictions for all physical quantities that appear in it. In this sense, one of the problems of the Standard Model (SM) is its large number of free parameters -which are 19 for the case of massless neutrinos- while for the massive case there are between 27 and 29 parameters, depending on the nature of neutrinos. Among these parameters are the  mixing angles, the CP-violating phases, the Higgs quartic coupling and the various Yukawa couplings. In the SM these couplings cannot be derived from first principles. Their values are obtained from experimental data and are evolved within the energy scale of validity of the SM using the renormalization group equations (RGEs).  A successful extension of the SM should be able to address, at least partly, the proliferation of free parameters in order to be predictive.

The idea of reduction of couplings (RoC) consists in obtaining \textit{renormalization group invariant} (RGI) relations, which become constraints for the parameters of the considered theory.  In this framework, we define as Finite Grand Unified Theories (FUTs)
those Supersymmetric Grand Unified Theories ($N = 1$ SUSY GUTs) whose $\beta$-functions
are vanishing. 
For this reason, all-loop FUTs, in particular  based, among others, on the RoC method,  are considered to have built in a powerful tool for dealing with the plethora of free parameters and yielding predictions.

In 1983, the dimensionless couplings of $N = 1$ SUSY GUTs were analyzed and
expressions known as the first and second finiteness conditions were
derived \cite{Parkes:1984dh}, which are one-loop RGI relations that ensure finiteness up
to two-loops. These conditions also demonstrate the fact that finiteness cannot be achieved within the class of models that contain   $U\left(1\right)$ as gauge group. As a result, one cannot pursue a finite version of the Minimal Supersymmetric Standard Model (MSSM), but has to look towards larger gauge groups. Studies on GUT gauge groups have determined the set of irreducible
representations (irreps) to which matter superfields should be assigned in
order to fulfill the first finiteness condition\cite{Hamidi:1984ft,Jiang:1988na,Rajpoot:1984zq}. 
In particular, the $SU\left(5\right)$ gauge group requires the following set of irreducible representations (irreps), 
 $\left(\textbf{5},{\bf{\overline{5}}},{\bf{10}},{\bf{\overline{10}}},{\bf{15}}, {\bf{\overline{15}}},{\bf{24}}\right)$ with corresponding multiplicity $\left(4,7,3,0,0,0,1\right)$, in
order to have a chance to become finite at one-loop and a phenomenologically
appealing model.

Given that FUTs seem to be necessarily at least $N = 1$ supersymmetric the
finiteness conditions should be extended to the dimensionful soft supersysymmetry
breaking sector, in addition to the dimensionless one.
Past works that employ RGI relations first on the dimensionless sector of
such theories have predicted the mass of the top quark \cite{Kapetanakis:1992vx,Mondragon:1993tw,Kubo:1994bj} and then
employing the dimensionful sector too, the Higgs boson \cite{Heinemeyer:2008qw} years
before their respective experimental discoveries.
Furthermore, the reduced and finite models studied in refs. \cite{Heinemeyer:2013nza,Heinemeyer:2018zpw,Heinemeyer:2020ftk} predict naturally heavy supersymmetric spectra. A detail study of their discovery potential in (near) future colliders can be found in ref. \cite{Heinemeyer:2020nzi}. It should be noted that the concept of finiteness has also been employed in cosmological models of inflation \cite{Elizalde:2015nya}. For comprehensive reviews on RoC, their application in finite unified theories and their predictions see refs. \cite{Sibold:1972258,Heinemeyer:2019vbc}.

The above-mentioned studies on the subject have focused on the third generation of quarks. In the present work, superpotentials with mixing terms among generations are considered and Yukawa couplings for all three generations are taken into account,  both in the dimensionless and the dimensionful sectors of the considered $SU(5)$ finite theories.
One of the main tests that each GUT has to face to be physically
acceptable is the agreement of its results with the experimental
data on proton decay \cite{Langacker:1980js,Kazakov:1995cy}. In this context, two fine tunings are performed in our models, one in the
superpotential and another in the soft SUSY breaking terms, in order to
ensure that the coloured triplets are heavier than the GUT scale, necessary for the solution to the doublet-triplet splitting problem and to avoid  fast proton decay.  

There have also been earlier studies of quark masses and mixings in finite models with three generations based on discrete symmetries,  mainly in $A_4$, $S_4$, and $Q_6$ \cite{Babu:2002in,Babu:2002ki,Jimenez:2014fqa}. These studies were mainly focused in finding examples of all-loop finite models. In the present work we also study $SU(5)$ finite theories addressing the flavour problem with discrete symmetries as $S_{3}$ and/or $Z_{n}$, but looking also for two-loop finiteness solutions, which, although parametric, give rise to a wider array of viable mass matrix textures.
%into account, among other improvements, the necessary conditions for the absence of proton decay.
The general form of the mass matrices compatible with finiteness was given in ref. \cite{Babu:2002in}. In here, we use these results as a basis for finding explicitly  these two-loop finiteness solutions, where the value of the mass entries is determined (up to phases) in terms of one  or two parameters only, whose values are restricted. 
In some cases these two-loop solutions can be promoted to all-loop finite ones with no free parameters except from the phases. We also determine the minimum number of phases allowed in the models.

This paper is organized as follows: In Section 2 we present a basic introduction to the \textit{reduction of couplings (RoC)} method, and in Section 3 the application of the method to render finite supersymmetric unified field theories. In Section 4 we present a general discussion on $N=1$ $SU(5)$ finite Grand Unified models mass matrices, including the soft breaking terms and the absence of proton decay.  As a first approach we present in section 5  a model with diagonal mass matrices and the resulting fit for the quark masses.  
In Section 6 we present a model based on the permutational symmetry $S_3$ supplemented with cyclic symmetries $Z_3\times Z_2$, which is finite to all-loops, but that leads to unrealistic mass matrices.
In Section 7 we present finite $SU(5)$ models that exhibit cyclic symmetries, where the up and down mass matrices share the same texture.  We find that most solutions are parametric, leading to two-loop finite models, some of which have viable mass textures. A particular solution for one of these models displays more symmetry and is indeed all-loop finite, and leads to realistic mass matrices. We include also a subsection on how to include the phases in the mass matrices in the least arbitrary possible way. Section 8 is dedicated to some concluding remarks.

\section{Theoretical Basis of Reduction of couplings}

We start with the core idea of the \textit{reduction of couplings} method. 
The goal is to express all parameters of a theory in terms of one basic parameter (which we will call `primary' coupling), using RGI relations among them. Considering such relations exist, they are in general of the form $\Phi (g_1,\cdots,g_A) ~=~\mbox{const.}$
and they should satisfy the  partial differential equation (PDE)
\begin{equation}
\mu\,\frac{d \Phi}{d \mu} = {\vec \nabla}\Phi\cdot {\vec \beta} ~=~
\sum_{a=1}^{A}
\,\beta_{a}\,\frac{\partial \Phi}{\partial g_{a}}~=~0~,
\end{equation}
where $\beta_a$ are the  $\beta$-functions of $g_a$.
Solving the above PDE is equivalent to solving the following set of ordinary differential equations (ODEs), which are called Reduction Equations (REs)
\cite{Zimmermann:1984sx,Oehme:1984yy,Oehme:1985jy},
\begin{equation}
\beta_{g} \,\frac{d g_{a}}{d g} =\beta_{a}~,~a=1,\cdots,A-1~,
\label{redeq}
\end{equation}
where now $g$ is the primary coupling and $\beta_g$ its respective $\beta$-function. The crucial demand is that the  REs admit power series solutions
\begin{equation}
g_{a} = \sum_{n}\rho_{a}^{(n)} \, g^{2n+1} ~,
\label{powerser}
\end{equation}
which preserve perturbative renormalizability. Without this requirement, we just
trade each ``dependent'' coupling with an integration constant. Remarkably, the uniqueness of such a solution can be already decided at the one-loop level
\cite{Zimmermann:1984sx,Oehme:1984yy,Oehme:1985jy}. However, the reduction
is usually not ``complete'', which means that not all of the couplings are reduced in favour
of the primary one, leading to the so called ``partial reduction''
\cite{Kubo:1985up,Kubo:1988zu}.

We proceed with the reduction scheme for parameters with mass dimension 1 and 2.
A  number of conditions are required (see for example
\cite{Piguet:1989pc}) and, following \cite{Kubo:1996js}, we can introduce
mass parameters and couplings carrying mass dimension
\cite{Breitenlohner:2001pp,Zimmermann:2001pq}
in the same way as dimensionless couplings.
Consider the superpotential
\begin{equation}
W= \frac{1}{2}\,\mu^{ij} \,\Phi_{i}\,\Phi_{j}+
\frac{1}{6}\,C^{ijk} \,\Phi_{i}\,\Phi_{j}\,\Phi_{k}~,
\label{supot}
\end{equation}
and the SSB sector Lagrangian
\begin{equation}
-{\cal L}_{\rm SSB} =
\frac{1}{6} \,h^{ijk}\,\phi_i \phi_j \phi_k
+
\frac{1}{2} \,b^{ij}\,\phi_i \phi_j
+
\frac{1}{2} \,(m^2)^{j}_{i}\,\phi^{*\,i} \phi_j+
\frac{1}{2} \,M\,\lambda_i \lambda_i+\mbox{h.c.},
\label{supot_l}
\end{equation}
where $\phi_i$'s are the scalar fields of the corresponding superfields $\Phi_i$'s and
$M$ is the gaugino mass which is denoted by $\lambda$. Some well-known and useful relations:
\begin{enumerate}[(i)]
\item The $\beta$-function of the gauge coupling at one-loop level is given by
\cite{Parkes:1984dh,West:1984dg,Jones:1985ay,Jones:1984cx,Parkes:1985hh}
\begin{equation}
\beta^{(1)}_{g}=\frac{d g}{d t} =
  \frac{g^3}{16\pi^2}\,\left[\,\sum_{i}\,T(R_{i})-3\,C_{2}(G)\,\right]~.
\label{betag}
\end{equation}
\item The anomalous dimension $\gamma^{(1)}\,^i_j$, at a one-loop level, of a chiral~superfield  is
\begin{equation}
\gamma^{(1)}\,^i_j=\frac{1}{32\pi^2}\,\left[\,
C^{ikl}\,C_{jkl}-2\,g^2\,C_{2}(R_{i})\delta^i_j\,\right]~.
\label{gamay}
\end{equation}
\item The $\beta$-functions of $C_{ijk}$'s, at one-loop level, following the
$N = 1$ non-renormali\-zation theorem
\cite{Wess:1973kz,Iliopoulos:1974zv,Fujikawa:1974ay},
are expressed in terms of the anomalous dimensions of the fields involved
\end{enumerate}
\beq
\beta_C^{ijk} =
  \frac{d C_{ijk}}{d t}~=~C_{ijl}\,\gamma^{l}_{k}+
  C_{ikl}\,\gamma^{l}_{j}+
  C_{jkl}\,\gamma^{l}_{i}~.
\label{betay}
\eeq
We proceed by assuming
that the REs admit~power series~solutions:
\beq
C^{ijk} = g\,\sum_{n=0}\,\rho^{ijk}_{(n)} g^{2n}~.
\label{Yg}
\eeq
Trying to obtain all-loop results we turn to relations among $\beta$-functions.
The spurion technique
\cite{Fujikawa:1974ay,Delbourgo:1974jg,Salam:1974pp,Grisaru:1979wc,Girardello:1981wz}
gives all-loop relations among SSB $\beta$-functions
\cite{Yamada:1994id,Kazakov:1997nf,Jack:1997pa,Hisano:1997ua,Jack:1997eh,Avdeev:1997vx,Kazakov:1998uj}.
Then, assuming that the reduction of $C^{ijk}$ is possible to all orders
\beq
\label{Cbeta}
\frac{dC^{ijk}}{dg} = \frac{\beta^{ijk}_C}{\beta_g}~,
\eeq
as well as for $h^{ijk}$
\beq
\label{h2NEW}
h^{ijk} = - M \frac{dC(g)^{ijk}}{d\ln g}~,
\eeq
then it can be proven
\cite{Jack:1999aj,Kobayashi:1998iaa}
that the following relations are all-loop RGI
\begin{align}
M &= M_0~\frac{\beta_g}{g} ,  \label{M-M0} \\
h^{ijk}&=-M_0~\beta_C^{ijk},  \label{hbeta}  \\
b^{ij}&=-M_0~\beta_{\mu}^{ij},\label{bij}\\
(m^2)^i_j&= \frac{1}{2}~|M_0|^2~\mu\frac{d\gamma^i{}_j}{d\mu}~,
\label{scalmass}
\end{align}
where $M_0$~is an~arbitrary reference mass scale to be specified
(note that in both assumptions we do not rely on specific solutions of these equations).\\
As a next step we substitute the last equation, Eq.(\ref{scalmass}), by a more general RGI sum rule that
holds to all orders \cite{Kobayashi:1997qx,Kobayashi:1998jq}
\begin{equation}
\begin{split}
m^2_i+m^2_j+m^2_k &=
|M|^2 \left\{~
\frac{1}{1-g^2 C_2(G)/(8\pi^2)}\frac{d \ln C^{ijk}}{d \ln g}
+\frac{1}{2}\frac{d^2 \ln C^{ijk}}{d (\ln g)^2}~\right\}\\
& \qquad\qquad +\sum_l
\frac{m^2_l T(R_l)}{C_2(G)-8\pi^2/g^2}
\frac{d \ln C^{ijk}}{d \ln g}~,
\label{sum2}
\end{split}
\end{equation}
which leads to the following one-loop relation \cite{Kobayashi:1997qx}
\beq
m^2_i+m^2_j+m^2_k=|M|^2~.\label{1loopsumrule}
\eeq

\section{Finiteness in \emph{N} = 1 Supersymmetric Theories}

Consider an $N=1$ globally supersymmetric gauge theory, which is chiral and anomaly free,
where $G$ is the gauge group and $g$ its associated gauge coupling. The theory has the
superpotential of Eq.(\ref{supot}), while the one-loop gauge and $C_{ijk}$'s $\beta$-function
are given by Eq.(\ref{betag}) and Eq.(\ref{betay}) respectively and the one-loop anomalous dimensions
of the chiral superfields by Eq.(\ref{gamay}).\\
Demanding the vanishing of all one-loop $\beta$-functions, Eqs.(\ref{betag},\ref{gamay}) lead to
the relations
\begin{align}
\sum _i T(R_{i})& = 3 C_2(G) \,,
\label{1st}     \\
 C^{ikl} C_{jkl} &= 2\delta ^i_j g^2  C_2(R_i)~.
\label{2nd}
\end{align}
The finiteness conditions for an $N=1$ supersymmetric theory with $SU(N)$
associated group is found in \cite{Rajpoot:1984zq}, while discussion of the
no-charge renormalization and anomaly free requirements can be found in  \cite{Rajpoot:1985aq}.
It should be noted that conditions (\ref{1st}) and (\ref{2nd}) are necessary and sufficient to
ensure finiteness at the two-loop level
\cite{Parkes:1984dh,West:1984dg,Jones:1985ay,Jones:1984cx,Parkes:1985hh}.

The requirement of finiteness, at the one-loop level, in softly broken SUSY theories
demands additional constraints among the soft terms of the SSB sector \cite{Jones:1984cu},
while, once more, these one-loop requirements assure two-loop finiteness, too \cite{Jack:1994kd}.
These conditions impose restrictions on the irreducible~representations $R_i$ of the gauge group
$G$ as well as on the Yukawa couplings. For example, since $U(1)$s are not compatible with condition
(\ref{1st}), the MSSM is excluded. Therefore, a GUT is initially required, with the MSSM being its low energy theory.

The non trivial point is that the relations among couplings (gauge and Yukawa) which are imposed
by the conditions (\ref{1st}) and (\ref{2nd}) should hold at any energy scale.
The necessary and sufficient~condition is~to
require that~such relations are~solutions to~the  REs (see \refeq{Cbeta})
\beq
\beta _g
\frac{d C_{ijk}}{dg} = \beta _{ijk} ~,
\label{redeq2}
\eeq
holding at all orders. We note, once more, that the existence of one-loop level power series solution
guarantees the all-order series.

There exists the following theorem
\cite{Lucchesi:1987he,Lucchesi:1987ef}
which points down which are the the necessary and sufficient conditions
in order for an $N=1$ SUSY theory to be all-loop finite. 
In refs
\cite{Lucchesi:1987he,Lucchesi:1987ef, Piguet:1986td,Piguet:1986pk,Ensign:1987wy,Lucchesi:1996ir} it was shown that for an $N=1$ SUSY Yang-Mills~theory, based on a simple gauge group,
if the following four conditions are fulfilled:\\
(i) No gauge anomaly is present.\\
(ii) The $\beta$-function of the gauge coupling is zero at one-loop level
\beq
\beta^{(1)}_g = 0 =\sum_i T(R_{i})-3\,C_{2}(G).
\eeq
(iii) The condition of vanishing for the one-loop anomalous dimensions of matter fields,
\beq
  \gamma^{(1)}{}_{j}^{i}~=~0
  =\frac{1}{32\pi^2}~[ ~
  C^{ikl}\,C_{jkl}-2~g^2~C_{2}(R)\delta_j^i ],
\eeq
admits solution of the form
\beq
C_{ijk}=\rho_{ijk}g,~\qquad \rho_{ijk}\in \mathbb{C}~.
\label{soltheo}
\eeq
(iv) When considered as solutions of vanishing Yukawa~$\beta$-functions
(at one-loop order), i.e. $\beta_{ijk}=0$, the above solutions are isolated and non-degenerate.\\
Then,~each of~the solutions in Eq.(\ref{soltheo})~can be extended uniquely to a formal power series in $g$,
and the~associated super Yang-Mills models~depend on the~single coupling constant $g$ with a vanishing, at all orders, $\beta$-function.

A crucial development was done in Ref.\cite{Kazakov:1997nf}, where it was shown that it is possible to choose certain RGI surfaces, so as to reach all-loop finiteness.  This result is based on the all-loop relations among the $\beta$-functions of the soft supersymmetry breaking terms and those of the rigid supersymmetric theory, noticing that on certain RGI surfaces the partial differential operators  acting on the $\beta$- and $\gamma$-functions of the rigid theory can be
transformed into total derivatives. This way the all-loop finiteness of the
$\beta$- and $\gamma$-functions of the rigid theory can be transferred into the $\beta$-functions of the soft supersymmetry breaking terms. Thus, a completely all-loop finite $N = 1$ SUSY gauge theory can be constructed, including the soft supersymmetry breaking terms.  For a detailed discussion of this procedure and a complete set of references see \cite{Mondragon:2021kec}.

\section{Finite \texorpdfstring{$SU(5)$}{TEXT} models with three generations}

As mentioned in the introduction, a classification of finite models with different unified gauge groups can be found in \cite{Hamidi:1984ft}. To allow for three generations of fermions and the correct breaking from $SU(5)$ to the SM group, the following content is singled out $({\bf{{5}}},{\bf{\bar 5}},{\bf{10}},{\bf{24}})$ with corresponding multiplicities $(4,7,3,1)$. We need exactly 3 copies of the irreps ${\bf{\bar{5}}}$ and ${\bf{10}}$ to accommodate the three known generations of fermions and 1 in the adjoint to break the gauge group to the MSSM one.
In this section we describe the general setup with this matter content. Reviews of previous work on FUTs based on $SU(5)$ and $SU(3)^3$ and their predictions, only considering the third generation, can be found in \cite{Sibold:1972258, Heinemeyer:2019vbc}.

\subsection{General SU(5) FUT superpotential and the second finiteness condition \label{GeneralSU5FUTSuperpotential}}

For the above-mentioned matter content we will use the following notation:
3 ${\bar{\Psi}}_{a^\prime i}$ superfields in the  $\bf{\bar{5}}$ irrep characterise the down-type antiquarks, charged leptons and neutrinos,
 3 $X_{a^\prime}^{ij}$ superfields in the  $\bf{10}$ represent the up-type quarks and antiquarks, and the down-type quarks and the charged antileptons,
4 $\mathcal{H}_a^i$ and 4 $\bar{\mathcal{H}}_{ai}$ in the $\bf{5}$ and  $\bf{\bar{5}}$ irrep respectively are assigned to the Higgs fields, and $\Sigma_{\ j}^i$ is  in the adjoint representation $\bf{24}$. 
Throughout the present work the indices $i,j,k,\ \ldots $ will be used for the gauge group representation, the primed indices $a^\prime,b^\prime,c^\prime,\ \ldots$ for the three fermion generations and the non-primed $a,b,c,\ldots$ label the four Higgs fields $\mathcal{H}_{a}$ and ${\bar{\mathcal{H}}}_{a}$.

The most general superpotential  with the above mentioned content, consistent with preserved R-parity is 

\begin{dmath}
	{\mathcal{W}_{SU(5)-R}}=  {\bar{g}}_{a^\prime b^\prime a}{\bar{\Psi}}_{b^\prime i}X_{a^\prime}^{ij}{\bar{\mathcal{H}}}_{aj}+\frac{1}{2}g_{a^\prime b^\prime a}\epsilon_{ijklm}X_{a^\prime}^{ij}X_{b^\prime}^{kl}\mathcal{H}_a^m+f_{ab}{\bar{\mathcal{H}}}_{ai}{\Sigma}_{\ j}^{i}\mathcal{H}_b^j+\frac{1}{3!}p{\Sigma}_{\ j}^{i}\ {\Sigma}_{\ k}^{j}\ {\Sigma}_{\ i}^{k}+\frac{1}{2}\lambda^{\left(\Sigma\right)}{\Sigma}_{\ j}^{i}{\Sigma}_{\ i}^{j}+m_{ab}{\bar{\mathcal{H}}}_{ai}\mathcal{H}_b^i \textrm{   .}
	\label{SuperpotencialGeneralParidadR}
\end{dmath}

Imposing the {\it second finiteness condition}, i.e. taking the anomalous dimensions equal to zero (\ref{2nd}), to the most general superpotential the following system of equations arises \cite{Kapetanakis:1992vx}:

\begin{displaymath}
\gamma_{\mathcal{\bar{H}}}: 4\bar{g}_{ija}\bar{g}^{ijb}+\frac{24}{5}f_{ac}f^{bc}+4q_{iac}q^{ibc}=\frac{24}{5}g^{2}\delta_{a}^{b} \textrm{ ,}
\end{displaymath}
\begin{displaymath}
\gamma_{\mathcal{H}}: 3g_{ija}g^{ijb}+\frac{24}{5}f_{ca}f^{cb}+\frac{24}{5}h_{ia}h^{ib}= \frac{24}{5}g^{2}\delta_{a}^{b} \textrm{ ,}
\end{displaymath}
\begin{displaymath}
\gamma_{\bar{\Psi}}: 4\bar{g}_{kia}\bar{g}^{kja}+\frac{24}{5}h_{ia}h^{ja}+4g^\prime_{ikl}{g^\prime}^{jkl}=\frac{24}{5}g^{2}\delta_{a}^{b} \textrm{ ,}
\end{displaymath}
\begin{displaymath}
\gamma_{X}: 2\bar{g}_{ika}\bar{g}^{jka}+3g_{ika}g^{jka}+q_{iab}q^{jab}+g^\prime_{kli}{g^\prime}^{klj}=\frac{36}{5}g^{2}\delta_{i}^{j} \textrm{ ,}
\end{displaymath}
\begin{equation}
\gamma_{\Sigma} : f_{ab}f^{ab}+\frac{21}{5}pp^{*}+h_{ia}h^{ia}=10g^{2} \textrm{ .}
\label{SistemaEcuacionesGeneral}
\end{equation}
Notice that in the case R-parity is preserved some of the couplings in (\ref{SistemaEcuacionesGeneral}), namely $h, g'$ and $q$, will not appear.
In order to be able to obtain an all-loop finite theory, isolated and non-degenerate solutions to such a system are required, which implies that the superpotenial has more symmetry.

\subsection{General Higgs sector rotation of \texorpdfstring{$SU(5)$}{TEXT} FUTs, doublet-triplet splitting and fine tuning}

    Once the $SU(5)$ gauge symmetry is broken, it will be assumed that the resulting theory is the  MSSM \cite{Heinemeyer:2020ftk,Kubo:1994xa,Kubo:1995hm,Heinemeyer:2010xt,Heinemeyer:2021zrk}. To this effect,  rotation transformations $(R,S)$ are performed on the Higgs sector $\left(\mathcal{H}_a,{\bar{\mathcal{H}}}_a\right)$ of the original FUT theory, to the MSSM basis $\left(\mathcal{K}_a,{\bar{\mathcal{K}}}_a\right)$, where only the pair $\mathcal{K}_3$ and ${\bar{\mathcal{K}}}_3$ are expected to acquire non-zero vacuum expectation values  \cite{Jones:1984qd,Leon:1985jm,Hamidi:1984gd,Kapetanakis:1992es,Mondragon:1993tw,Yoshioka:1997yt}. These rotations $\left(R,S\right)$ are generated from orthogonal and tetradimensional matrices, since they act on the space of the four pairs of Higgs fields. The rotation matrices $R$ and $S$ are applied as:

\begin{equation}
R_{ac}\mathcal{H}_c^j=\mathcal{K}_a^j \textrm{   ,}
\quad\quad {\bar{\mathcal{H}}}_{ci}S_{ca}={\bar{\mathcal{K}}}_{ai} \textrm{   .}
\label{HbSKb}
\end{equation}
In addition, a fine tuning must be implemented to ensure that the part associated with the doublet of the pair $(\mathcal{K}_3,{\bar{\mathcal{K}}}_3)$ is the only light one to match those of the MSSM, while the rest of the doublets and triplets in the sector $\left(\mathcal{K}_a,{\bar{\mathcal{K}}}_a\right)$ are heavy. The superpotential sector associated with doublet-triplet splitting can be written in a simplified form as:
\begin{equation}
\mathcal{W}_{DTS}={\bar{\mathcal{H}}}^{\left(3\right)}M^{\left(3\right)}\mathcal{H}^{\left(3\right)}+{\bar{\mathcal{H}}}^{\left(2\right)}M^{\left(2\right)}\mathcal{H}^{\left(2\right)}  \textrm{   ,}
\label{WDTS}
\end{equation}
where the mass matrices of the doublets and triplets are:
\begin{equation}
M^{\left(2\right)}=-3\omega f+m   \textrm{   ,}
\label{M2}
\end{equation}
\begin{equation}
M^{\left(3\right)}=2\omega f+m   \textrm{   .}
\label{M3}
\end{equation}
The fine tuning consists of assigning a value to the elements of the matrix $m$, such that after the rotation we are assured that the doublets belonging to the third Higgs pair $(\mathcal{K}_{3}^{(2)},\bar{\mathcal{K}}_{3}^{(2)})$ are light and there is no early proton decay.

The inverse of the rotation transformations (\ref{HbSKb}) are substituted on the Higgs fields of the superpotential sector associated with the doublet-triplet splitting
\begin{dmath}
	\mathcal{W}_{DTS}={\bar{\mathcal{K}}}^{\left(3\right)}S^T\left(2\omega f+m\right)R^T\mathcal{K}^{\left(3\right)}+{\bar{\mathcal{K}}}^{\left(2\right)}S^T\left(-3\omega f+m\right)R^T\mathcal{K}^{\left(2\right)}
	={\bar{\mathcal{K}}}^{\left(3\right)}\hat{M}^{(3)}\mathcal{K}^{\left(3\right)}+{\bar{\mathcal{K}}}^{\left(2\right)}\hat{M}^{(2)}\mathcal{K}^{\left(2\right)}
	\textrm{   .}
\end{dmath}
The rotation matrices transform $M^{\left(2\right)}$ to a diagonal form similar to the following matrix ${\hat{M}}^{\left(2\right)}$:

\begin{equation}
{\hat{M}}^{\left(2\right)}=\left(\begin{matrix}\omega&0&0&0\\0&\omega&0&0\\0&0&\omega^\prime&0\\0&0&0&\omega\\\end{matrix}\right)   \textrm{   ,}
\label{PrimerAjusteFinoMh2}
\end{equation}

\noindent where $\omega\approx M_{GUT}$ and $|\omega^\prime|\approx M_Z$. Therefore, one can find another way to express $M^{\left(2\right)}$ by ${\hat{M}}^{\left(2\right)}$ as

\begin{equation}
M^{\left(2\right)}=S{\hat{M}}^{\left(2\right)}R   \textrm{   .}
\label{M2S1S2Mh2R2R1}
\end{equation}
From this last relation (\ref{M2S1S2Mh2R2R1}), together with the definition (\ref{M2}) of $M^{\left(2\right)}$ coming from the superpotential, one can derive the matrix $m$,

\begin{displaymath}
 M^{\left(2\right)}=S{\hat{M}}^{\left(2\right)}R=-3\omega f+m   \textrm{   ,}
\end{displaymath}
\begin{equation}
 m=S{\hat{M}}^{\left(2\right)}R+3\omega f   \textrm{   .}
\label{mMatriz}
\end{equation}
The matrix $M^{\left(3\right)}$, defined in (\ref{M3}), is rewritten through rotations as:

\begin{equation}
M^{\left(3\right)}=2\omega f+m=S{\hat{M}}^{\left(2\right)}R+5\omega f   \textrm{   ,}
\end{equation}
\noindent while ${\hat{M}}^{\left(3\right)}$ is:

\begin{equation}
{\hat{M}}^{\left(3\right)}=S^TM^{\left(3\right)}R^T=S^T\left(2\omega f+m\right)R^T={\hat{M}}^{\left(2\right)}+5\omega S^TfR^T   \textrm{   .}
\label{Mh3}
\end{equation}    
To avoid early proton decay, the determinant of ${\hat{M}}^{\left(3\right)}$ must be greater than or close to $\omega^4$ \cite{Leon:1985jm}, i.e.

\begin{equation}
\det{\left[{\hat{M}}^{\left(3\right)}\right]}\ge\omega^4 \textrm{   .}
\label{detMh3}
\end{equation}
On the other hand, some of the soft SUSY breaking terms resemble those involved in the doublet-triplet splitting of the superpotential. Evidently, the process of doublet-triplet splitting in the superpotential is performed between superfields, while the doublet-triplet splitting for the soft-breaking terms is performed with the component fields. The terms that resemble the doublet-triplet splitting of the superpotential are:
\begin{equation}
-{\bar{H}}_{aj}{h_f}_{ab}{\phi_{\Sigma}}_{\ i}^jH_b^i-{\bar{H}}_{ai}b_{m_{ab}}H_b^i   \textrm{   .}
\end{equation}
Taking the vacuum expectation value of the field $\phi_{\Sigma}$

\begin{dmath}
	-{\bar{H}}_{aj}{h_f}_{ab}\left\langle{\phi_{\Sigma}}_{\ i}^j\right\rangle H_b^i-{\bar{H}}_{ai}b_{m_{ab}}H_b^i=-{\bar{H}}_a^{\left(3\right)}\left(2\omega{h_f}_{ab}+b_{m_{ab}}\right)H_b^{\left(3\right)}-{\bar{H}}_a^{\left(2\right)}\left(-3\omega{h_f}_{ab}+b_{m_{ab}}\right)H_b^{\left(2\right)}   \textrm{   ,}
	\label{DTSSSB1}
\end{dmath}

\noindent and using the solutions (\ref{hbeta}), coming from the dimensional sector for the $h$ type couplings, equation (\ref{DTSSSB1}) becomes
\begin{equation}
=-{\bar{H}}_a^{\left(3\right)}\left(-2\omega M f_{ab}+b_{m_{ab}}\right)H_b^{\left(3\right)}-{\bar{H}}_a^{\left(2\right)}\left(3\omega M f_{ab}+b_{m_{ab}}\right)H_b^{\left(2\right)}   \textrm{   .}
\end{equation}
The second term of this expression generates a contribution to the squared mass of the light doublets. To prevent this term from affecting the result of the previous fine tuning, i.e. causing the pair of doublets that are expected to be light to become heavy, a fine tuning has to be performed also on the parameters $b_{m_{ab}}$. The simplest way to solve this second fine tuning problem is to relate it to the first one. In this respect, the ansatz suggested by the first fine tuning and the finiteness solution to the $h_{f_{ab}}$ couplings is useful. Thus, if $h_{f_{ab}}$ is related to $f_{ab}$, then $b_{m_{ab}}$ can be related to $m_{ab}$. The form of this tuning is \cite{Leon:1985jm,Yoshioka:1997yt}

\begin{equation}
	b_{m_{ab}}\approx-Mm_{ab}   \textrm{   .}
\label{bmabAjusteFinoLSSB}
\end{equation}
This expression should not be surprising, since some SUGRA models propose it as well as other relations among parameters, valid at Planck scale. Moreover, several of these relations are extrapolated to the GUT scale, assuming that it is a good approximation. 
The effect of substituting the fine tuning of $b_{m_{ab}}$ in the soft doublet-triplet splitting is:
\begin{dmath}
	-{\bar{H}}_a^{\left(3\right)}\left(-2\omega M f_{ab}+b_{m_{ab}}\right)H_b^{\left(3\right)}-{\bar{H}}_a^{\left(2\right)}\left(3\omega M f_{ab}+b_{m_{ab}}\right)H_b^{\left(2\right)}\approx-{\bar{H}}_a^{\left(3\right)}\left(-2\omega M f_{ab}-Mm_{ab}\right)H_b^{\left(3\right)}-{\bar{H}}_a^{\left(2\right)}\left(3\omega M f_{ab}-Mm_{ab}\right)H_b^{\left(2\right)}=M{\bar{H}}_a^{\left(3\right)}\left(2\omega f_{ab}+m_{ab}\right)H_b^{\left(3\right)}-M{\bar{H}}_a^{\left(2\right)}\left(3\omega f_{ab}-m_{ab}\right)H_b^{\left(2\right)}   \textrm{   .}
\end{dmath}
By applying the inverse of the rotation transformations (\ref{HbSKb}) to the corresponding scalar component fields 
\begin{equation}
\mathcal{H}_b^j=\left(R^T\right)_{bc}\mathcal{K}_c^j\ \ \Rightarrow\ \ \ H=R^TK   \textrm{   ,}
\end{equation}
\begin{equation}
{\bar{\mathcal{H}}}_{ai}={\bar{\mathcal{K}}}_{ei}\left(S^T\right)_{ea}\ \ \Rightarrow\ \ \ \bar{H}=\bar{K}S^T   \textrm{   ,}
\end{equation}
and remembering the expressions (\ref{M2}), (\ref{M3}), (\ref{M2S1S2Mh2R2R1}) and (\ref{Mh3}), the second doublet-triplet splitting reduces to
\begin{align}
-{\bar{H}}_{aj}{h_f}_{ab}\left\langle{\phi_{\Sigma}}_{\ i}^j\right\rangle H_b^i-{\bar{H}}_{ai}b_{m_{ab}}H_b^i\approx M{\bar{H}}_a^{\left(3\right)}M_{ab}^{\left(3\right)}H_b^{\left(3\right)}+M{\bar{H}}_a^{\left(2\right)}M_{ab}^{\left(2\right)}H_b^{\left(2\right)} \nonumber \\
=M{\bar{K}}_e^{\left(3\right)}\left(S^T\right)_{ea}M_{ab}^{\left(3\right)}\left(R^T\right)_{bd}K_d^{\left(3\right)}+M{\bar{K}}_e^{\left(2\right)}\left(S^T\right)_{ea}M_{ab}^{\left(2\right)}\left(R^T\right)_{bd}K_d^{\left(2\right)} \nonumber \\
=M{\bar{K}}_a^{\left(3\right)}{\hat{M}}_{ab}^{\left(3\right)}K_b^{\left(3\right)}+M{\bar{K}}_a^{\left(2\right)}{\hat{M}}_{ab}^{\left(2\right)}K_b^{\left(2\right)}~.
\end{align}

\subsection{General mass matrices of \texorpdfstring{$SU(5)$}{TEXT} FUT \label{MatricesMasaGenerales}}

Fermion mass matrices are defined in supersymmetry by the expression
\begin{equation}
M=\left.\left\langle\frac{\partial^2\mathcal{W}\left(\Phi_1,\ldots,\Phi_n\right)}{\partial\Phi_i\partial\Phi_j}\right\rangle\right|_{\Phi_i=\phi_i,\Phi_j=\phi_j} \textrm{   ,}
\label{ExpresionMatrizDeMasa}
\end{equation}
where $\phi_{i}$ and $\phi_{j}$ are the scalar component fields associated with the superfields $\Phi_{i}$ and $\Phi_{j}$.
In the present case, the mass matrices for up and down type quarks will be of the general form:
\begin{equation}
	M_u=\left(\begin{matrix}{g_{11a}\left\langle\mathcal{H}_a^5\right\rangle}&{g_{12a}\left\langle\mathcal{H}_a^5\right\rangle}&{g_{13a}\left\langle\mathcal{H}_a^5\right\rangle}\\ {g_{21a}\left\langle\mathcal{H}_a^5\right\rangle}&{g_{22a}\left\langle\mathcal{H}_a^5\right\rangle}&{g_{23a}\left\langle\mathcal{H}_a^5\right\rangle}\\ {g_{31a}\left\langle\mathcal{H}_a^5\right\rangle}&{g_{32a}\left\langle\mathcal{H}_a^5\right\rangle}&{g_{33a}\left\langle\mathcal{H}_a^5\right\rangle}\\\end{matrix}\right) \textrm{   ,}
	\label{MuGeneral}
\end{equation}

\begin{equation}
	M_d=\left(\begin{matrix}{{\bar{g}}_{11a}\left\langle{\bar{\mathcal{H}}}_{a5}\right\rangle}&{{\bar{g}}_{12a}\left\langle{\bar{\mathcal{H}}}_{a5}\right\rangle}&{{\bar{g}}_{13a}\left\langle{\bar{\mathcal{H}}}_{a5}\right\rangle}\\{{\bar{g}}_{21a}\left\langle{\bar{\mathcal{H}}}_{a5}\right\rangle}&{{\bar{g}}_{22a}\left\langle{\bar{\mathcal{H}}}_{a5}\right\rangle}&{{\bar{g}}_{23a}\left\langle{\bar{\mathcal{H}}}_{a5}\right\rangle}\\{{\bar{g}}_{31a}\left\langle{\bar{\mathcal{H}}}_{a5}\right\rangle}&{{\bar{g}}_{32a}\left\langle{\bar{\mathcal{H}}}_{a5}\right\rangle}&{{\bar{g}}_{33a}\left\langle{\bar{\mathcal{H}}}_{a5}\right\rangle}\\\end{matrix}\right) \textrm{   .}
    \label{MdGeneral}
\end{equation}

As already stated, after the breaking of the $SU(5)$ group we will have the MSSM, which has only one pair of Higgs doublets. This implies a rotation of the Higgs sector to the MSSM basis,  such that only one Higgs is coupled to the up type particles and another one to the down type ones, and a fine tuning is performed to ensure that only two pairs of Higgs doublets are light. We can find expressions for the MSSM parameters in terms of the rotation matrices $R$   and $S$  (\ref{HbSKb}) defined in the previous sections.  
The relations between parameters of the MSSM and $SU(5)$ FUT at the GUT scale are shown below:\\

{\underline{Bilinear couplings}:
\begin{equation}
\mu=\left(S^T\right)_{3a}\left(-3\omega f_{ab}+m_{ab}\right)\left(R^T\right)_{b3}
\label{ParametroMu} ~,~~
B_\mu=\left(S^T\right)_{3a}\left({b_m}_{ab}-3\omega{h_f}_{ab}\right)\left(R^T\right)_{b3} ~.
\end{equation}

{\underline{Trilinear couplings $A$ of the soft breaking lagrangian}}:
\begin{equation}
A_{u,a^\prime b^\prime}={h_g}_{a^\prime b^\prime a}\left(R^T\right)_{a3}={h_g}_{a^\prime b^\prime a}R_{3a} ~,
\end{equation}
\begin{equation}
A_{d,a^\prime b^\prime}={h_{\bar{g}}}_{a^\prime b^\prime a}\left(S^T\right)_{3a}={h_{\bar{g}}}_{a^\prime b^\prime a}S_{a3}
 \textrm{   ,   }\\
A_{e,a^\prime b^\prime}={h_{\bar{g}}}_{a^\prime b^\prime a}\left(S^T\right)_{3a}={h_{\bar{g}}}_{a^\prime b^\prime a}S_{a3} ~.
\end{equation}

{\underline{Soft breaking scalar masses}}:
\begin{equation}
m_{H_d}^2=\left(S^T\right)_{3a}\left(m_{\bar{H}}^2\right)_{ab}S_{b3} ~,~~~
m_{H_u}^2=R_{3a}\left(m_{H}^2\right)_{ab}\left(R^T\right)_{b3} ~,
\end{equation}
\begin{equation}
m_{Q,a^\prime b^\prime}^2=\left(m_{\widetilde{\chi}}^2\right)_{a^\prime b^\prime} ~,~~
m_{L,a^\prime b^\prime}^2=\left(m_{\widetilde{\bar{\psi}}}^2\right)_{a^\prime b^\prime} ~.
\end{equation}
\begin{equation}
m_{u,a^\prime b^\prime}^2=\left(m_{\widetilde{\chi}}^2\right)_{a^\prime b^\prime} ~,~~
m_{d,a^\prime b^\prime}^2=\left(m_{\widetilde{\bar{\psi}}}^2\right)_{a^\prime b^\prime}~,~~
m_{e,a^\prime\ b^\prime}^2=\left(m_{\widetilde{\chi}}^2\right)_{a^\prime b^\prime} ~.
\end{equation}

{\underline{Gaugino mass parameters $M_{i}$}}:
\begin{equation}
M_1=M_2=M_3=M ~.
\end{equation}

\section{First approach: Diagonal model FUTA}

The FUTA model represents a first approach to the finiteness discussion, in which the superpotential is constructed so that the mass matrices are diagonal, i.e., no mixing terms appear between generations \cite{Kapetanakis:1992vx}. We can first treat the problem under this approximation, since the elements of the diagonal of the CKM matrix are the ones with the highest value, to exemplify some features of three generation FUT models. 
The superpotential of the FUTA model, written in the convention shown in the subsection \ref{GeneralSU5FUTSuperpotential}, is:

\begin{align}
\mathcal{W}_{SU\left(5\right)}=&\sum_{a^\prime=1}^{3}\left[{\bar{g}}_{a^\prime a^\prime a^\prime}{\bar{\Psi}}_{a^\prime i}X_{a^\prime}^{ij}{\bar{\mathcal{H}}}_{a^\prime j}+\frac{1}{2}g_{a^\prime a^\prime a^\prime}\epsilon_{ijklm}X_{a^\prime}^{ij}X_{a^\prime}^{kl}\mathcal{H}_{a^\prime}^m\right] \nonumber \\
&+f_{44}{\bar{\mathcal{H}}}_{4j}\Sigma_{\ i}^j\mathcal{H}_4^i+\frac{1}{3!}p\Sigma_{\ j}^i\ \Sigma_{\ k}^j\ \Sigma_{\ i}^k+\mathcal{W}_m \textrm{ ,}
\label{SuperpotencialFUTA}
\end{align}
where the term $\mathcal{W}_m$ represents the bilinear terms of the superpotential, while the rest of the terms are trilinear.
Application of the system of equations (\ref{SistemaEcuacionesGeneral}) yields the following unification relations between gauge and Yukawa couplings:
\begin{displaymath}
\left|g_{111}\right|^2=\left|g_{222}\right|^2=\left|g_{333}\right|^2=\frac{8}{5}g^2\textrm{,}\ \ \left|{\bar{g}}_{111}\right|^2=\left|{\bar{g}}_{222}\right|^2=\left|{\bar{g}}_{333}\right|^2=\frac{6}{5}g^2\textrm{,}\ \ 
\end{displaymath}
\begin{displaymath}
\left|g_{234}\right|^2=0\textrm{,}\ \ \left|{\bar{g}}_{234}\right|^2=\left|{\bar{g}}_{324}\right|^2=0\textrm{,}\ \ \ \  \left|f_{11}\right|^2=\left|f_{22}\right|^2=\left|f_{33}\right|^2=0\textrm{,} \ \ \ \left|f_{44}\right|^2=g^2 \textrm{,}
\end{displaymath}
\begin{equation}
\left|p\right|^2=\frac{15}{7}g^2 \textrm{   .} \label{Yuk_diag}
\end{equation}
To achieve the above, the superpotential (\ref{SuperpotencialFUTA}) follows a $Z_7\times Z_3\times Z_2$ symmetry, where the charges for the superfields are shown in Table \ref{TablaSimetriasCiclicasFUTA} \cite{Heinemeyer:2010xt}:
\begin{table}[H]
	\begin{center}
		\resizebox{14cm}{!}{
			\begin{tabular}{|c|c|c|c|c|c|c|c|c|c|c|c|c|c|c|c|}
				\hline
				& & & & & & & & & & & & & & &
				\\
				$Z_{n}$ & $\bar{\Psi}_{1}$ & $\bar{\Psi}_{2}$ & $\bar{\Psi}_{3}$ & $X_{1}$ & $X_{2}$ & $X_{3}$ & $\mathcal{H}_{1}$ & $\mathcal{H}_{2}$ & $\mathcal{H}_{3}$ & $\mathcal{H}_{4}$ & $\mathcal{\bar{H}}_{1}$ & $\mathcal{\bar{H}}_{2}$ & $\mathcal{\bar{H}}_{3}$ & $\mathcal{\bar{H}}_{4}$ & $\Sigma$ \\
				\hline
				$Z_{7}$ & $4$ & $1$ & $2$ & $1$ & $2$ & $4$ & $5$ & $3$ & $6$ & $0$ & $-5$ & $-3$ & $-6$ & $0$ & $0$ \\
				\hline
				$Z_{3}$ & $0$ & $0$ & $0$ & $1$ & $2$ & $0$ & $1$ & $2$ & $0$ & $0$ & $-1$ & $-2$ & $0$ & $0$ & $0$ \\
				\hline
				$Z_{2}$ & $1$ & $1$ & $1$ & $1$ & $1$ & $1$ & $0$ & $0$ & $0$ & $0$ & $0$ & $0$ & $0$ & $0$ & $0$ \\
				\hline
			\end{tabular}
		}
		{	\caption{\label{TablaSimetriasCiclicasFUTA} Cyclic discrete symmetries of the diagonal FUTA model.
		}}
	\end{center}
\end{table}
\noindent This model was studied using six parameters associated with the Higgs sector rotation matrices. The theoretical quark masses are given by the product of the  Yukawa coupling given by the finiteness conditions Equation~(\ref{Yuk_diag}), the corresponding Higgs vev (either $v_u$ or $v_d$), and the rotation angles (see the Appendix) . Then, a $\chi^2$ fit was performed to find the theoretical value of the quark masses at $M_Z$, as well as $\tan\beta$.  The fit was performed by 
comparing with seven known observables, which were $m_u\left(M_Z\right)$, $m_c\left(M_Z\right)$, $m_t\left(M_Z\right)$, $m_d\left(M_Z\right)$, $m_s\left(M_Z\right)$, $m_b\left(M_Z\right)$, and we use the value of $m_\tau\left(M_Z\right)$ to find $\tan\beta$. 
The $\chi^2$ function was defined as:
\bea
    \nn \chi_{r}^{2} &=& \frac{1}{\nu}\sum_{k=1}^{7}{\frac{(m_{k}^{Th}(M_{Z})-m_{k}^{Exp}(M_{Z}))^{2}}{\Delta m_{k}^{2}}} \\ \nn
    &=& \frac{(m_{u}(M_{Z})^{Th}-m_{u}^{Exp}(M_{Z}))^{2}}{\Delta m_{u}^{2}}+\frac{(m_{d}(M_{Z})^{Th}-m_{d}^{Exp}(M_{Z}))^{2}}{\Delta m_{d}^{2}}\\ \nn
   && +\frac{(m_{c}(M_{Z})^{Th}-m_{c}^{Exp}(M_{Z}))^{2}}{\Delta m_{c}^{2}}+\frac{(m_{s}(M_{Z})^{Th}-m_{s}^{Exp}(M_{Z}))^{2}}{\Delta m_{s}^{2}}\\ \nn
  && +\frac{(m_{t}(M_{Z})^{Th}-m_{t}^{Exp}(M_{Z}))^{2}}{\Delta m_{t}^{2}}
   +\frac{(m_{b}(M_{Z})^{Th}-m_{b}^{Exp}(M_{Z}))^{2}}{\Delta m_{b}^{2}}\\
   &&+\frac{(m_{\tau}(M_{Z})^{Th}-m_{\tau}^{Exp}(M_{Z}))^{2}}{\Delta m_{\tau}^{2}}
\eea
where $\Delta m_{k}$ is the experimental uncertainty associated to the corresponding mass.

The masses and the obtained $\tan{\beta}$ for the minimum of $\chi_r^2$ are given in Table \ref{FUTAResults}.

\begin{table}[H]
\begin{center}
	\resizebox{14cm}{!}{
\begin{tabular}{|c|c|c|c|c|c|c|c|c|}
	\hline
	& & & & & & & & \\
$m_u\left(M_Z\right)$ & $m_c\left(M_Z\right)$ & $m_t(M_Z)$ & $m_d\left(M_Z\right)$ & $m_s\left(M_Z\right)$ & $m_b\left(M_Z\right)$ & $m_\tau\left(M_Z\right)$ & $\tan{\beta}$ & $\chi_{r_{min}}^2$ \\
	& & & & & & & & \\
\hline
& & & & & & & & \\
$0.0012GeV$ & $0.626GeV$ & $171.8GeV$ & $0.00278GeV$ & $0.0595GeV$ & $2.86GeV$ & $1.74623GeV$ & $57.4$ & $0.152$\\
& & & & & & & & \\
\hline
\end{tabular}} \caption{\label{FUTAResults} Results of the quark and $\tau$ lepton masses and $\tan\beta$ corresponding to the minimum of $\chi_{r}^{2}$ in FUTA.}
\end{center}
\end{table}

\noindent This calculation was done assuming a 20\% of threshold corrections to the bottom mass, coming from SUSY breaking  corrections at low energies and from other threshold corrections at the GUT scale. Finally, to estimate proton decay, the determinant of the mass matrix associated with the Higgs triplets ${\hat{M}}^{\left(3\right)}$ must satisfy the relation (\ref{detMh3}). The fourth root of this determinant is:
\begin{equation}
(\det[\hat{M}^{(3)}])^{\frac{1}{4}}=1.125\omega \textrm{   .}
\end{equation}
This result agrees with heavy triplets  larger than the GUT scale in all Higgs fields belonging to the fundamental and
antifundamental representations, a necessary requirement to avoid fast proton decay \cite{Nath:2006ut,Babu:2013jba}.

\section{An \texorpdfstring{$SU(5)$}{TEXT} FUT model based on \texorpdfstring{$S_{3}$}{TEXT}}

In this section we present an $SU(5)$ FUT model based on the
discrete flavour group $S_3$.  This is the smallest non-Abelian
discrete group, which represents the permutation of three objects, or
equivalently, the rotations and reflections that leave invariant an
equilateral triangle.  It has three irreducible representations
(irreps): a doublet $\bf 2$, a symmetric singlet $\bf 1_S$ and an
antisymmetric singlet $\mathbf{{1}_{A}}$ (see for instance \cite{Ishimori:2010au}).
Its tensor products are given by
\begin{equation}
(\psi_{1})_{\mathbf{{1}_{S}}}\otimes(\varphi_{1})_{\mathbf{{1}_{A}}}=(\psi_{1}\varphi_{1})_{\mathbf{{1}_{A}}} \textrm{ ,}
\end{equation}
\begin{equation}
(\psi_{1})_{\mathbf{{1}_{A}}}\otimes(\varphi_{1})_{\mathbf{{1}_{A}}}=(\psi_{1}\varphi_{1})_{\mathbf{{1}_{S}}} \textrm{ ,}
\end{equation}
\begin{equation}
\psi_{\mathbf{2}}\otimes (\varphi_{1})_{\mathbf{{1}_{S}}}=\left(\begin{array}{c}
\psi_{1} \\
\psi_{2} \\
\end{array}\right)_{\mathbf{2}}\otimes (\varphi_{1})_{\mathbf{{1}_{S}}}=\left(\begin{array}{c}
\psi_{1} \varphi_{1} \\
\psi_{2} \varphi_{1} \\
\end{array}\right)_{\mathbf{2}} \textrm{ ,}
\end{equation}
\begin{equation}
\psi_{\mathbf{2}}\otimes (\varphi_{1})_{\mathbf{{1}_{A}}}=\left(\begin{array}{c}
\psi_{1} \\
\psi_{2} \\
\end{array}\right)_{\mathbf{2}}\otimes (\varphi_{1})_{\mathbf{{1}_{A}}}=\left(\begin{array}{c}
-\psi_{2}\varphi_{1} \\
\psi_{1}\varphi_{1} \\
\end{array}\right)_{\mathbf{2}} \textrm{ ,}
\end{equation}
\begin{equation}
\resizebox{0.9\textwidth}{!}{$\psi_{\mathbf{2}} \otimes \varphi_{\mathbf{2}}=\left(\begin{array}{c}
\psi_{1} \\
\psi_{2} \\
\end{array}\right)_{\mathbf{2}} \otimes \left(\begin{array}{c}
\varphi_{1} \\
\varphi_{2} \\
\end{array}\right)_{\mathbf{2}}=\left(\psi_{1}\varphi_{1} +
\psi_{2}\varphi_{2}\right)_{\mathbf{{1}_{S}}}\oplus
\left(\psi_{1}\varphi_{2}-\psi_{2}\varphi_{1}\right)_{\mathbf{{1}_{A}}} \oplus
\left(\begin{array}{c}
\psi_{2}\varphi_{2}-\psi_{1}\varphi_{1} \\
\psi_{1}\varphi_{2}+\psi_{2}\varphi_{1} \\
\end{array}\right)_{\mathbf{2}}$} \textrm{ .}
\end{equation}
At low energies, $S_3$ has proven to be very successful in accommodating the masses and mixing in the quark sector \cite{Kubo:2003iw, GonzalezCanales:2013pdx}, as well as in the leptonic sector \cite{Mondragon:2007af}, in this latter case predicting correctly two of the mixing angles in some models \cite{GonzalezCanales:2012blg}. The assignments of the fields to the $S_3$ irreps can be seen in Table \ref{tab:S3_SU5}. For the quarks and leptons, we assign the third family to the symmetric singlet of $S_3$, and the first two to the doublet.  With respect to the Higgs fields and their conjugates,  we assign two to the doublet irrep and one to the singlet, and the fourth Higgs is assigned to the antisymmetric singlet. The $SU(5)$ field in the adjoint is also assigned to the symmetric singlet.

\begin{table}[H]
    \centering
    \resizebox{\textwidth}{!}{ 
    \begin{tabular}{|c|c|c|c|c|c|c|c|c|c|c|c|}
    \hline
     & & & & & & & & & & & \\
     Superfields & $\left(\begin{array}{c}
         \bar{\Psi}_{1i} \\
         \bar{\Psi}_{2i} \\
     \end{array}\right)$ & $\bar{\Psi}_{3i}$ &  $\left(\begin{array}{c}
         X_{1}^{ij} \\
         X_{2}^{ij} \\
     \end{array}\right)$ & $X_{3}^{ij}$ & $\left(\begin{array}{c}
         \mathcal{H}_{1}^{i} \\
         \mathcal{H}_{2}^{i} \\
     \end{array}\right)$ & $\mathcal{H}_{3}^{i}$ & $\mathcal{H}_{4}^{i}$ & $\left(\begin{array}{c}
         \bar{\mathcal{H}}_{1i} \\
         \bar{\mathcal{H}}_{2i} \\
     \end{array}\right)$ & $\bar{\mathcal{H}}_{3i}$ & $\bar{\mathcal{H}}_{4i}$ & $\Sigma_{\ j}^{i}$ \\
     &  & & & & & & & & & &\\
      \hline
    Irreducible & & & & & & & & & & & \\
    representations of & $\bf{2}$ & $\bf{1_{S}}$ & $\bf{2}$ & $\bf{1_{S}}$ & $\bf{2}$ & $\bf{1_{S}}$ & $\bf{1_{A}}$ & $\bf{2}$ & $\bf{1_{S}}$ & $\bf{1_{A}}$ & $\bf{1_{S}}$ \\
    $S_{3}$ &  & & & & & & & & & &\\
    \hline
    \end{tabular}
    }
    \caption{Assignement of the superfields to the $S_3$ irreducible representations.}
    \label{tab:S3_SU5}
\end{table}

\noindent The $SU(5)$ superpotential, respecting the $S_3$ symmetry with the above assignemnts and symmetric under R-parity, is given by

\begin{dmath}
\mathcal{W}_{{SU\left(5\right)\times S}_3}={\bar{g}}_{121}\left({\bar{\mathcal{H}}}_{1j}X_1^{ij}{\bar{\Psi}}_{2i}+{\bar{\mathcal{H}}}_{1j}X_2^{ij}{\bar{\Psi}}_{1i}+{\bar{\mathcal{H}}}_{2j}X_1^{ij}{\bar{\Psi}}_{1i}-{\bar{\mathcal{H}}}_{2j}X_2^{ij}{\bar{\Psi}}_{2i}\right)+{\bar{g}}_{113}\left({\bar{\mathcal{H}}}_{3j}X_1^{ij}{\bar{\Psi}}_{1i}+{\bar{\mathcal{H}}}_{3j}X_2^{ij}{\bar{\Psi}}_{2i}\right)+{\bar{g}}_{124}\left({\bar{\mathcal{H}}}_{4j}X_1^{ij}{\bar{\Psi}}_{2i}-{\bar{\mathcal{H}}}_{4j}X_2^{ij}{\bar{\Psi}}_{1i}\right)+{\bar{g}}_{311}\left({\bar{\mathcal{H}}}_{1j}X_3^{ij}{\bar{\Psi}}_{1i}+{\bar{\mathcal{H}}}_{2j}X_3^{ij}{\bar{\Psi}}_{2i}\right)+{\bar{g}}_{131}\left({\bar{\mathcal{H}}}_{1j}X_1^{ij}{\bar{\Psi}}_{3i}+{\bar{\mathcal{H}}}_{2j}X_2^{ij}{\bar{\Psi}}_{3i}\right)+{\bar{g}}_{333}{\bar{\mathcal{H}}}_{3j}X_3^{ij}{\bar{\Psi}}_{3i}+\frac{1}{2}g_{121}\epsilon_{ijklm}\left(2\mathcal{H}_1^mX_1^{ij}X_2^{kl}+\mathcal{H}_2^mX_1^{ij}X_1^{kl}-\mathcal{H}_2^mX_2^{ij}X_2^{kl}\right)+\frac{1}{2}g_{113}\epsilon_{ijklm}\left(\mathcal{H}_3^mX_1^{ij}X_1^{kl}+\mathcal{H}_3^mX_2^{ij}X_2^{kl}\right)+g_{131}\epsilon_{ijklm}\left(\mathcal{H}_1^mX_1^{ij}X_3^{kl}+\mathcal{H}_2^mX_2^{ij}X_3^{kl}\right)+\frac{1}{2}g_{333}\epsilon_{ijklm}\mathcal{H}_3^mX_3^{ij}X_3^{kl}+f_{11}\left({\bar{\mathcal{H}}}_{1j}\mathcal{H}_1^i\Sigma_{\ i}^j+{\bar{\mathcal{H}}}_{2j}\mathcal{H}_2^i\Sigma_{\ i}^j\right)+f_{33}{\bar{\mathcal{H}}}_{3j}\mathcal{H}_3^i\Sigma_{\ i}^j+f_{44}{\bar{\mathcal{H}}}_{4j}\mathcal{H}_4^i\Sigma_{\ i}^j+\frac{1}{3!}p\Sigma_{\ j}^i\Sigma_{\ k}^j\Sigma_{\ i}^k+m_{11}\left({\bar{\mathcal{H}}}_{1i}\mathcal{H}_1^i+{\bar{\mathcal{H}}}_{2i}\mathcal{H}_2^i\right)+m_{33}{\bar{\mathcal{H}}}_{3i}\mathcal{H}_3^i+m_{44}{\bar{\mathcal{H}}}_{4i}\mathcal{H}_4^i+\frac{1}{2}\lambda^{\left(\Sigma\right)}\Sigma_{\ j}^i\Sigma_{\ i}^j \textrm{   .}
\label{SuperpotencialGeneralS3Paridad_R}
\end{dmath}
This $SU(5)\times S_3$ superpotential leads to the following mass matrices for up and down type quarks:
\begin{equation}
M_u=\left(\begin{matrix}g_{121}\left\langle\mathcal{H}_2^5\right\rangle+g_{113}\left\langle\mathcal{H}_3^5\right\rangle&g_{121}\left\langle\mathcal{H}_1^5\right\rangle&g_{131}\left\langle\mathcal{H}_1^5\right\rangle\\g_{121}\left\langle\mathcal{H}_1^5\right\rangle&-g_{121}\left\langle\mathcal{H}_2^5\right\rangle+g_{113}\left\langle\mathcal{H}_3^5\right\rangle&g_{131}\left\langle\mathcal{H}_2^5\right\rangle\\g_{131}\left\langle\mathcal{H}_1^5\right\rangle&g_{131}\left\langle\mathcal{H}_2^5\right\rangle&g_{333}\left\langle\mathcal{H}_3^5\right\rangle\\\end{matrix}\right) \textrm{   ,}
\end{equation}
\begin{equation}
M_d=\left(\begin{matrix}{\bar{g}}_{121}\left\langle{\bar{\mathcal{H}}}_{25}\right\rangle+{\bar{g}}_{113}\left\langle{\bar{\mathcal{H}}}_{35}\right\rangle&{\bar{g}}_{121}\left\langle{\bar{\mathcal{H}}}_{15}\right\rangle+{\bar{g}}_{124}\left\langle{\bar{\mathcal{H}}}_{45}\right\rangle&{\bar{g}}_{131}\left\langle{\bar{\mathcal{H}}}_{15}\right\rangle\\{\bar{g}}_{121}\left\langle{\bar{\mathcal{H}}}_{15}\right\rangle-{\bar{g}}_{124}\left\langle{\bar{\mathcal{H}}}_{45}\right\rangle&-{\bar{g}}_{121}\left\langle{\bar{\mathcal{H}}}_{25}\right\rangle+{\bar{g}}_{113}\left\langle{\bar{\mathcal{H}}}_{35}\right\rangle&{\bar{g}}_{131}\left\langle{\bar{\mathcal{H}}}_{25}\right\rangle\\{\bar{g}}_{311}\left\langle{\bar{\mathcal{H}}}_{15}\right\rangle&{\bar{g}}_{311}\left\langle{\bar{\mathcal{H}}}_{25}\right\rangle&{\bar{g}}_{333}\left\langle{\bar{\mathcal{H}}}_{35}\right\rangle\\ \end{matrix}\right) \textrm{   .}
\end{equation}
Extensions of the SM with $S_3$ as flavour symmetry and three Higgs
doublets, without supersymmetry, have been studied in the literature
and they lead to the form of the mass matrices shown here.
Here we explore an all-loop finite model, which has as flavour
symmetry $S_3$ and extra cyclic symmetries.
The $M_d$ mass matrix exhibits a similar structure to the one in 
\cite{GonzalezCanales:2013pdx}, for a non-supersymmetric 4 Higgs
doublet model with $S_3$ symmetry. The up type mass matrix $M_u$ shows
a different structure, since in the model presented here the up quark
information comes from the $X_{a^\prime}$ superfields, that commute
among themselves, and the symmetrization of $S_3$ introduces a minus
sign in the antisymmetric singlet representation $\mathbf{{1}_{A}}$
assigned to $\mathcal{H}_4$, leading to the cancellation of the
following terms $\frac{1}{2}\left(g_{124}X_1X_2\mathcal{H}_4+g_{214}X_2X_1\mathcal{H}_4\right)=\frac{1}{2}g_{124}\left(X_1X_2\mathcal{H}_4-X_2X_1\mathcal{H}_4\right)=0$. 

\begin{sloppypar}
The finiteness solutions  (\ref{SistemaEcuacionesGeneral}) to the
general superpotential (\ref{SuperpotencialGeneralS3Paridad_R}) lead to
parametric solutions to the squared norm of the trilineal couplings,
$\left|f_{11}\right|^2=\left|f_{22}\right|^2=\frac{1}{4}\left(4g_5^2-5\left|g_{131}\right|^2\right)$
and
$\left|f_{33}\right|^2=\frac{1}{2}\left(-4g_5^2\
  +5\left|g_{131}\right|^2\right)$ for the couplings $f_{ab}$  of the
three Higgs fields coupled to the fermions. Since the square norm is
positive definite 
$\left|f_{aa}\right|^2\geq0$, the only possible solution is
$\left|f_{11}\right|^2=\left|f_{22}\right|^2=\left|f_{33}\right|^2=0$. 
This  imposes an additional restriction to the system of equations, 
$\left|g_{333}\right|^2=\frac{2}{5}\left(4g_5^2-5\left|g_{113}\right|^2\right)$
and 
$\left|{\bar{g}}_{333}\right|^2=-\frac{3}{10}\ \left(4\ g_5^2-5\
  \left|g_{113}\right|^2\right)$, which implies these couplings are
also zero, 
i.e. $\left|g_{333}\right|^2=\left|{\bar{g}}_{333}\right|^2=0$.
These conditions lead to the proposed $S_3$ model, which is the only
finite model allowed under the irrep assignement we considered, and that
allows for all-loop finiteness solutions, which in turn imply
additional cyclic symmetries, as we show in the next section.
\end{sloppypar}

\subsection{Model 1: Finite \texorpdfstring{$S_{3}$}{TEXT} to all orders}

Here we present the above-mentioned all-loop finite $SU(5)\times S_3$
model, which also exhibits the cyclic symmetries shown in 
 Table \ref {TablaSimetriasCiclicasModelS3T2II}. The superpotential is given by

\begin{table}[t]
	\begin{center}
		\resizebox{14cm}{!}{
			\begin{tabular}{|c|c|c|c|c|c|c|c|c|c|c|c|c|c|c|c|}
				\hline
				& & & & & & & & & & & & & & &
				\\
				$Z_{n}$ & $\bar{\Psi}_{1}$ & $\bar{\Psi}_{2}$ & $\bar{\Psi}_{3}$ & $X_{1}$ & $X_{2}$ & $X_{3}$ & $\mathcal{H}_{1}$ & $\mathcal{H}_{2}$ & $\mathcal{H}_{3}$ & $\mathcal{H}_{4}$ & $\mathcal{\bar{H}}_{1}$ & $\mathcal{\bar{H}}_{2}$ & $\mathcal{\bar{H}}_{3}$ & $\mathcal{\bar{H}}_{4}$ & $\Sigma$ \\
				\hline
				$Z_{2}$ & $1$ & $1$ & $1$ & $1$ & $1$ & $1$ & $0$ & $0$ & $0$ & $0$ & $0$ & $0$ & $0$ & $0$ & $0$ \\
				\hline
				$Z_{3}$ & $0$ & $0$ & $1$ & $1$ & $1$ & $2$ & $0$ & $0$ & $1$ & $0$ & $1$ & $1$ & $2$ & $0$ & $0$ \\
                \hline
			\end{tabular}
		}
		{	\caption{\label{TablaSimetriasCiclicasModelS3T2II}Cyclic discrete symmetries of the model based on $S_{3}$. These symmetries allow us to obtain isolated and non-degenerate solutions to the system of equations (\ref{SistemaEcuacionesGeneral}).
		}}
	\end{center}
\end{table}

\begin{align}
&\mathcal{W}_{{SU\left(5\right)\times S}_3\times Z_2\times Z_3} ={\bar{g}}_{113}\left({\bar{\mathcal{H}}}_{3j}X_1^{ij}{\bar{\Psi}}_{1i}+{\bar{\mathcal{H}}}_{3j}X_2^{ij}{\bar{\Psi}}_{2i}\right)+{\bar{g}}_{311}\left({\bar{\mathcal{H}}}_{1j}X_3^{ij}{\bar{\Psi}}_{1i}+{\bar{\mathcal{H}}}_{2j}X_3^{ij}{\bar{\Psi}}_{2i}\right) \nonumber \\
& +{\bar{g}}_{131}\left({\bar{\mathcal{H}}}_{1j}X_1^{ij}{\bar{\Psi}}_{3i}+{\bar{\mathcal{H}}}_{2j}X_2^{ij}{\bar{\Psi}}_{3i}\right)+\frac{1}{2}g_{113}\epsilon_{ijklm}\left(\mathcal{H}_3^mX_1^{ij}X_1^{kl}+\mathcal{H}_3^mX_2^{ij}X_2^{kl}\right) \nonumber \\ 
& +g_{131}\epsilon_{ijklm}\left(\mathcal{H}_1^mX_1^{ij}X_3^{kl}+\mathcal{H}_2^mX_2^{ij}X_3^{kl}\right)+f_{44}{\bar{\mathcal{H}}}_{4i}\Sigma_{\ j}^i\mathcal{H}_4^j+\frac{1}{3!}p\Sigma_{\ j}^i\Sigma_{\ k}^j\Sigma_{\ i}^k \nonumber \\
&+m_{ab}{\bar{\mathcal{H}}}_a\mathcal{H}_b+\frac{1}{2}\lambda^{\left(\Sigma\right)}\Sigma {_{\ j}^{i}}\Sigma{_{\ i}^{j}} \textrm{   ,}
\label{SuperpotencialModeloS3}
\end{align}
which  has terms that mix the three generations,
contrary to FUTA which is diagonal.  On the other hand, the terms 
$m_{ab}{\bar{\mathcal{H}}}_a\mathcal{H}_b$ break the cyclic
symmetries, and are necessary to solve the doublet-triplet splitting.
These terms do not affect the finiteness conditions, which depend only
on  the trilinear terms.
The solutions to the finiteness conditions  are: 
\begin{displaymath}
	\left|g_{113}\right|^2=\frac{4}{5}g_5^2 \ \ \textrm{,} \ \ \left|g_{131}\right|^2=\frac{4}{5}g_5^2 \ \ \textrm{,} \ \ \left|{\bar{g}}_{113}\right|^2=\frac{3}{5}g_5^2\ \ \ ,\ \ \ \left|{\bar{g}}_{131}\right|^2=\frac{3}{5}g_5^2\ \ \ ,\ \ \ \left|{\bar{g}}_{311}\right|^2=\frac{3}{5}g_5^2 \textrm{   ,}
\end{displaymath}
\begin{equation}
	\left|f_{11}\right|^2=\left|f_{33}\right|^2=0\ \ \ ,\ \ \ \left|f_{44}\right|^2=g_5^2\ \ \ ,\ \ \ \left|p\right|^2=\frac{15}{7}g_5^2 \textrm{   },
\end{equation}
and they are isolated and non-degenerate, ensuring finiteness to all orders of perturbation theory. In general, the Yukawa couplings are complex, but since the finiteness
conditions involve only their squared norm this information is lost.
To maintain generality we have included them in the following
expressions, noting that they are in principle free parameters.
The solutions are:
\begin{displaymath}
    g_{113}=\frac{2}{\sqrt5}g_5e^{i\phi_1}\ \ \ ,\ \ \ g_{131}=\frac{2}{\sqrt5}g_5e^{i\phi_2} \textrm{   ,}
\end{displaymath}
\begin{equation}
    {\bar{g}}_{113}=\sqrt{\frac{3}{5}}g_5e^{i{\bar{\phi}}_1}\ \ \ ,\ \ \ {\bar{g}}_{131}=\sqrt{\frac{3}{5}}g_5e^{i{\bar{\phi}}_2}\ \ \ ,\ \ \ {\bar{g}}_{311}=\sqrt{\frac{3}{5}}g_5e^{i{\bar{\phi}}_3} \textrm{   ,}
\end{equation}
and $f_{11}$, $f_{22} = f_{33} = g_{333} = {\bar{g}}_{333} = 0$.  We
have also taken $f_{44}=g_5$ and $p=\sqrt{\frac{15}{7}}g_5$ as real,
to avoid extra sources of CP violation.

In all the FUT models previously studied, as well as in the models here presented,  the two-loop correction to the soft breaking sector sum rule (\ref{sum2}) vanishes. In the present case the one-loop sum rule (\ref{1loopsumrule}) leads to 11 equations that depend on 5 parameters, as shown below:
\begin{displaymath}
\resizebox{1\textwidth}{!}{$m_{{\widetilde{\bar{\psi}}}_1}^2=m_{{\widetilde{\bar{\psi}}}_2}^2=\frac{1}{2}\left(MM^\dag-2m_{{\bar{H}}_3}^2+m_{H_3}^2\right)~,~~
m_{{\widetilde{\bar{\psi}}}_3}^2=\frac{1}{2}\left(MM^\dag-2m_{{\bar{H}}_3}^2-2m_{H_2}^2+3m_{H_3}^2\right) , ~$}
\end{displaymath}
\begin{displaymath}
m_{{\widetilde{\chi}}_1}^2=m_{{\widetilde{\chi}}_2}^2=\frac{1}{2}\left(MM^\dag-m_{H_3}^2\right) ~, ~~
m_{{\widetilde{\chi}}_3}^2=\frac{1}{2}\left(MM^\dag-2m_{H_2}^2+m_{H_3}^2\right) ~,
\end{displaymath}
\begin{displaymath}
m_{H_1}^2=m_{H_2}^2 ~,~~
m_{{\bar{H}}_1}^2=m_{{\bar{H}}_2}^2=m_{{\bar{H}}_3}^2+m_{H_2}^2-m_{H_3}^2 ~,~~
m_{{\bar{H}}_4}^2=\frac{2}{3}MM^\dag-m_{H_4}^2 ~,~
\end{displaymath}
\beq
m_{\phi_\Sigma}^2=\frac{1}{3}MM^\dag \ \  .~
\label{SumRuleModelS3}
\eeq

\subsubsection{Mass matrices}
The up and down mass matrices for this model are:
\begin{equation}
M_u=\left(\begin{matrix}g_{113}\left\langle\mathcal{H}_3^5\right\rangle&0&g_{131}\left\langle\mathcal{H}_1^5\right\rangle\\0&g_{113}\left\langle\mathcal{H}_3^5\right\rangle&g_{131}\left\langle\mathcal{H}_2^5\right\rangle\\g_{131}\left\langle\mathcal{H}_1^5\right\rangle&g_{131}\left\langle\mathcal{H}_2^5\right\rangle&0\\\end{matrix}\right) \textrm{   ,}
\end{equation}
\begin{equation}
M_d=\left(\begin{matrix}{\bar{g}}_{113}\left\langle{\bar{\mathcal{H}}}_{35}\right\rangle&0&{\bar{g}}_{131}\left\langle{\bar{\mathcal{H}}}_{15}\right\rangle\\0&{\bar{g}}_{113}\left\langle{\bar{\mathcal{H}}}_{35}\right\rangle&{\bar{g}}_{131}\left\langle{\bar{\mathcal{H}}}_{25}\right\rangle\\{\bar{g}}_{311}\left\langle{\bar{\mathcal{H}}}_{15}\right\rangle&{\bar{g}}_{311}\left\langle{\bar{\mathcal{H}}}_{25}\right\rangle&0\\\end{matrix}\right) \textrm{   .}
\end{equation}
Unfortunately, it is clear from the textures that these will not render
a realistic spectrum at low energies, as two of the masses will be almost
degenerate.  The requirement of the flavour symmetry and finiteness 
constrained the parameters too much, leading to unrealistic mass
matrices.  This can be overcome by looking for different symmetries
and/or two-loop solutions. In the next section we  search for other finite
models,  based instead only on cyclic symmetries.

\section{\texorpdfstring{$SU(5)$}{TEXT} FUT models based on cyclic symmetries}

In this section we present examples of $SU(5)$ finite models including all three families, where
the finiteness requirement implies cyclic symmetries.  These models are similar to
the previously studied models FUTA and FUTB (see for instance ref. \cite{Heinemeyer:2011kd}), where the finiteness
conditions were only applied to the third generation.  We find 
the solutions to the finiteness conditions for each model,  and we show explicitly the sum rule for the soft scalars for one all-loop finite model.
Each solution leads to a particular texture
at the GUT scale.

Mass matrices are in general complex to account for CP violation, and
their phases add to the theory's free parameters.  As we already
mentioned, information
about these phases cannot be obtained via the finiteness conditions
eq(\ref{SistemaEcuacionesGeneral}), since these conditions only give 
information about the trilinear couplings norm squared, and thus these
phases are in principle free parameters.  A proposed solution in
ref.~\cite{Babu:2002ki} was to implement the phases through complex
vev's.  Here, we follow a different strategy, based on
refs.\cite{Kusenko:1993wu,Kusenko:1994zm}.  The idea consists in
quantifying the minimum number of independent phases, and then address
the question of where they are placed in the Yukawa matrices, by
performing phase invariant products. This way, we attempt to make a
deeper analysis of the RoC in FUTs, by including
the minimum number of phases in the Yukawa matrices in a
mathematically consistent way, in order to characterise the Yukawa
couplings both by their magnitude as well as their phases.

Ref.\cite{Babu:2002in} offers a classification of eight textures that
mass matrices may take to fulfill the requirement of cancellation of
the off diagonal terms of the anomalous dimensions
(\ref{SistemaEcuacionesGeneral}).  This classification is particularly
useful when considering FUT models that express cyclic
symmetries. This classification is divided into two block of four
matrices $V_3^{\left(i\right)}$, in which the quarks couple only to
three Higgs pairs$\left(\mathcal{H}_a,{\bar{\mathcal{H}}}_a\right)$,
and another four matrices $V_4^{\left(i\right)}$, where they couple to
the four Higgs pairs.  The subscripts $3$ and $4$ in
$V_3^{\left(i\right)}$ and $V_4^{\left(i\right)}$ denote the number of
Higgs coupled, whereas the superscript $\left(i\right)$, with
$i=1,\ldots,4$, classifies the four different matrices in each block.
For the case of three pairs of Higgs coupled to the fermions, the matrices are:

\begin{displaymath}
\resizebox{1\textwidth}{!}{$V_3^{\left(1\right)}=\left(\begin{matrix}g_{111}\left\langle\mathcal{H}_1^5\right\rangle&g_{123}\left\langle\mathcal{H}_3^5\right\rangle&g_{132}\left\langle\mathcal{H}_2^5\right\rangle\\g_{213}\left\langle\mathcal{H}_3^5\right\rangle&g_{222}\left\langle\mathcal{H}_2^5\right\rangle&g_{231}\left\langle\mathcal{H}_1^5\right\rangle\\g_{312}\left\langle\mathcal{H}_2^5\right\rangle&g_{321}\left\langle\mathcal{H}_1^5\right\rangle&g_{333}\left\langle\mathcal{H}_3^5\right\rangle\\\end{matrix}\right)\ \ ,\ \ V_3^{\left(2\right)}=\left(\begin{matrix}g_{112}\left\langle\mathcal{H}_2^5\right\rangle&g_{121}\left\langle\mathcal{H}_1^5\right\rangle&0\\g_{211}\left\langle\mathcal{H}_1^5\right\rangle&g_{223}\left\langle\mathcal{H}_3^5\right\rangle&g_{232}\left\langle\mathcal{H}_2^5\right\rangle\\0&g_{322}\left\langle\mathcal{H}_2^5\right\rangle&g_{333}\left\langle\mathcal{H}_3^5\right\rangle\\\end{matrix}\right) \ ,$}
\end{displaymath}
\begin{equation}
\resizebox{1\textwidth}{!}{$V_3^{\left(3\right)}=\left(\begin{matrix}g_{113}\left\langle\mathcal{H}_3^5\right\rangle&g_{121}\left\langle\mathcal{H}_1^5\right\rangle&0\\g_{211}\left\langle\mathcal{H}_1^5\right\rangle&g_{223}\left\langle\mathcal{H}_3^5\right\rangle&g_{232}\left\langle\mathcal{H}_2^5\right\rangle\\0&g_{322}\left\langle\mathcal{H}_2^5\right\rangle&g_{333}\left\langle\mathcal{H}_3^5\right\rangle\\\end{matrix}\right)\ \ ,\ \ V_3^{\left(4\right)}=\left(\begin{matrix}g_{111}\left\langle\mathcal{H}_1^5\right\rangle&0&0\\0&g_{223}\left\langle\mathcal{H}_3^5\right\rangle&g_{232}\left\langle\mathcal{H}_2^5\right\rangle\\0&g_{322}\left\langle\mathcal{H}_2^5\right\rangle&g_{333}\left\langle\mathcal{H}_3^5\right\rangle\\\end{matrix}\right) \ ,$}
\end{equation}
while for the case where the four pairs of Higgs couple to the fermions they are:
\begin{displaymath}
\resizebox{1\textwidth}{!}{$V_4^{\left(1\right)}=\left(\begin{matrix}g_{111}\left\langle\mathcal{H}_1^5\right\rangle&g_{124}\left\langle\mathcal{H}_4^5\right\rangle&g_{132}\left\langle\mathcal{H}_2^5\right\rangle\\g_{214}\left\langle\mathcal{H}_4^5\right\rangle&g_{222}\left\langle\mathcal{H}_2^5\right\rangle&g_{231}\left\langle\mathcal{H}_1^5\right\rangle\\g_{312}\left\langle\mathcal{H}_2^5\right\rangle&g_{321}\left\langle\mathcal{H}_1^5\right\rangle&g_{333}\left\langle\mathcal{H}_3^5\right\rangle\\\end{matrix}\right)\ \ ,\ \ V_4^{\left(2\right)}=\left(\begin{matrix}g_{112}\left\langle\mathcal{H}_2^5\right\rangle&g_{121}\left\langle\mathcal{H}_1^5\right\rangle&0\\g_{211}\left\langle\mathcal{H}_1^5\right\rangle&g_{222}\left\langle\mathcal{H}_2^5\right\rangle&g_{234}\left\langle\mathcal{H}_4^5\right\rangle\\0&g_{324}\left\langle\mathcal{H}_4^5\right\rangle&g_{333}\left\langle\mathcal{H}_3^5\right\rangle\\\end{matrix}\right) \ ,$}
\end{displaymath}
\begin{equation}
\resizebox{1\textwidth}{!}{$V_4^{\left(3\right)}=\left(\begin{matrix}g_{113}\left\langle\mathcal{H}_3^5\right\rangle&g_{121}\left\langle\mathcal{H}_1^5\right\rangle&g_{132}\left\langle\mathcal{H}_2^5\right\rangle\\g_{211}\left\langle\mathcal{H}_1^5\right\rangle&g_{222}\left\langle\mathcal{H}_2^5\right\rangle&g_{234}\left\langle\mathcal{H}_4^5\right\rangle\\g_{312}\left\langle\mathcal{H}_2^5\right\rangle&g_{324}\left\langle\mathcal{H}_4^5\right\rangle&g_{333}\left\langle\mathcal{H}_3^5\right\rangle\\\end{matrix}\right)\ \ ,\ \ V_4^{\left(4\right)}=\left(\begin{matrix}g_{113}\left\langle\mathcal{H}_3^5\right\rangle&g_{121}\left\langle\mathcal{H}_1^5\right\rangle&g_{132}\left\langle\mathcal{H}_2^5\right\rangle\\g_{211}\left\langle\mathcal{H}_1^5\right\rangle&g_{223}\left\langle\mathcal{H}_3^5\right\rangle&g_{234}\left\langle\mathcal{H}_4^5\right\rangle\\g_{312}\left\langle\mathcal{H}_2^5\right\rangle&g_{324}\left\langle\mathcal{H}_4^5\right\rangle&g_{333}\left\langle\mathcal{H}_3^5\right\rangle\\\end{matrix}\right) \ .$}
\end{equation}
These matrices are constructed using the
$g_{a^\prime b^\prime a}$ couplings and the $\mathcal{H}_a$ Higgs
fields of $M_u$. The $M_d$ matrices are calculated in the same way,
substituting the corresponding ${\bar{g}}_{a^\prime b^\prime a}$ and
${\bar{\mathcal{H}}}_a$, and have the same structure. Other matrices that are obtained by exchanging the Higgs indices  fall under this classification.  As an example, in one of the models studied here based on
$V_4^{\left(2\right)}$, the following exchange was made 
$\left(\mathcal{H}_2,{\bar{\mathcal{H}}}_2\right)\leftrightarrow\left(\mathcal{H}_4,{\bar{\mathcal{H}}}_4\right)$.

In the present work we explore the possible finite solutions that emerge from
these eight matrices when $M_u$ and $M_d$ share the same texture and
present the results from the models that come out from
$V_3^{\left(2\right)}$, $V_4^{\left(1\right)}$ and
$V_4^{\left(2\right)}$. In the construction of these models, we
considered the solutions that have only one or two pairs of
$\left(\mathcal{H}_3,{\bar{\mathcal{H}}}_3\right)$ Higgs fields in the
matrix elements.  This is because we are looking for hierarchical
solutions where the terms associated to
$\left(\mathcal{H}_3,{\bar{\mathcal{H}}}_3\right)$ are significantly
larger than the rest.  Due to the symmetries that emerge from applying
the finiteness conditions (\ref{SistemaEcuacionesGeneral}), and the
requirement that the solutions are positive (since they are squared
norms), it is always possible that some of the matrix elements turn
out to be zero. We also discard solutions coming from
$V_3^{\left(1\right)}$ and some from $V_4^{\left(1\right)}$, which
lead to FUTA, which is diagonal, or FUTB, which already has a block
diagonal structure, and solutions coming from
$V_4^{\left(3\right)}$, which lead to matrices with zeroes in the
diagonal.

In the following section we present the proposed $SU(5)$ models,
classified with the $V_3^{\left(2\right)}$, $V_4^{\left(1\right)}$,
and $V_4^{\left(2\right)}$ textures and  satisfying the finiteness
conditions (\ref{SistemaEcuacionesGeneral}).  In most cases the
solutions are parametric, which implies finiteness only at
two-loops. An all-loop finite model was also found, as an enhanced
symmetry case of one of the two-loop models.

The obtained solutions are parametric and cannot be negative, which
gives inequality relations among the couplings. Taking the values in
the upper or lower limits of these inequalities renders some of the
other couplings zero, which may be interpreted as the action of a
symmetry. For this reason if a limiting case produces viable textures
we look for the possible associated symmetries. 

Nevertheless, we were not able to find these symmetries in all cases
(which does not mean they do not exist), so in each model we propose a
particular value for the parametric solution, as an example of the
emerging textures. In this section, we present the form of the mass
matrices for a particular value of the parameter for models 2 and 3.
For model 4, which
exhibits a more interesting texture, we present the whole solution for
a particular case, as well as an all-loop version of the model.

A full analysis of the promising models, with a scan over the
parameter defining the solutions and their phenomenological
possibilites, is beyond the scope of this paper but will be presented
elsewhere.

\subsection{Model 2: Finite \texorpdfstring{$\mathbf{{V}_{3}^{\left({2}\right)}}$}{TEXT} at 2-loops}

In this model, the initial structure for up and down mass matrices is
of the type $V_3^{\left(2\right)}$. Assuming a diagonal $f_{ab}$, the
finiteness solutions to (\ref{SistemaEcuacionesGeneral}) found are
parametric. After requiring that they are positive (since they are
squared norms), the only solution found for $g_{121}$ is
$\left|g_{121}\right|^2=\frac{4}{5}g_5^2$. Then, inserting this
requirement into the parametric solutions, we find the solutions that
comply with this requirement.  In particular, it turns out that
$\left|f_{22}\right|^2$ and $\left|f_{33}\right|^2$ have to be zero,
since both are the same quantity but with opposite sign, i.e.
$\left|f_{22}\right|^2=\frac{1}{8}\left(8g_5^2-5\left|g_{112}\right|^2-10\left|g_{232}\right|^2\right)=-\left|f_{33}\right|^2\geq0$.  This in turn gives the final solutions, in terms of 
$\left|g_{112}\right|^2$ and $\left|g_{223}\right|^2$, which are:
\begin{displaymath}
	\left|g_{121}\right|^2=\left|g_{211}\right|^2=\frac{4}{5}g_5^2\ ,\ \left|g_{232}\right|^2=\left|g_{322}\right|^2=\frac{1}{10}\left(8g_5^2-5\left|g_{112}\right|^2\right)\ ,
\end{displaymath}
\begin{displaymath}
\left|g_{333}\right|^2=\frac{1}{5}\left(8g_5^2-5\left|g_{223}\right|^2\right)\ , \ 	\left|{\bar{g}}_{112}\right|^2=\left|{\bar{g}}_{121}\right|^2=\frac{3}{20}\left(8g_5^2-5\left|g_{112}\right|^2\right)\ ,
\end{displaymath}
\begin{displaymath}
\left|{\bar{g}}_{211}\right|^2=\frac{3}{4}\left|g_{112}\right|^2\ ,\ \left|{\bar{g}}_{223}\right|^2=\left|{\bar{g}}_{232}\right|^2=\frac{3}{20}\left(4g_5^2-5\left|g_{223}\right|^2\right)\ ,\
\end{displaymath}
\begin{displaymath}
\left|{\bar{g}}_{322}\right|^2=-\frac{3}{20}\left(4g_5^2-5\left|g_{112}\right|^2-5\left|g_{223}\right|^2\right)\ ,\ \left|{\bar{g}}_{333}\right|^2=\frac{3}{20}\left(4g_5^2+5\left|g_{223}\right|^2\right)\ ,\ 	
\end{displaymath}
\begin{equation}
	\left|f_{11}\right|^2=\left|f_{22}\right|^2=\left|f_{33}\right|^2=0\ ,\ \left|f_{44}\right|^2=g_5^2\ ,\ \left|p\right|^2=\frac{15}{7}g_5^2\ .
	\label{ParametricSolModel2}
\end{equation}
To guarantee that $\left|g_{112}\right|^2$ and
$\left|g_{223}\right|^2$ are non-negative, the following inequalities must
hold:
\begin{equation}
	0\le\left|g_{223}\right|^2\le\frac{4}{5}g_5^2  \ \ \textrm{,}   \ \ \frac{1}{5}\left(4g_5^2-5\left|g_{223}\right|^2\right)\le\left|g_{112}\right|^2\le\frac{8}{5}g_5^2 ~.
 \label{eq:condV32}
\end{equation}
This implies that $\left|g_{223}\right|^2$ may be zero, which is
desirable, since then $M_u$ depends only on one parameter
$\left\langle\mathcal{H}_3^5\right\rangle$. In this case,
$\left|{\bar{g}}_{223}\right|^2\neq0$, and thus $M_d$ has a different
texture from $M_u$, since it has two entries with
$\left\langle{\bar{\mathcal{H}}}_{35}\right\rangle$.

Another option is to set $\left|{\bar{g}}_{223}\right|^2=0$, which
leads to $\left|{\bar{g}}_{232}\right|^2 = 0$, and then
$\left|g_{223}\right|^2$ takes its other limiting value,
$\left|g_{223}\right|^2=\frac{4}{5}g_5^2$. This gives two more zeroes
in $M_d$, but we do not analyse this option, since it seems less
phenomenologically motivated than the other, given that the mass
difference between the third and the other two generations is more
pronounced in the up sector.

Setting $|g_{223}|^2=0$ in (\ref{ParametricSolModel2}) leads to the following interval for the allowed values of 
$|g_{112}|^2$:
\begin{equation}
	\frac{4}{5}g_5^2\le\left|g_{112}\right|^2\le\frac{8}{5}g_5^2 ~.
	\label{IntervaloDosg112}
\end{equation}
Again, taking the upper or lower end values in
(\ref{IntervaloDosg112}) may lead to more zeroes.  For instance, when
$\left|g_{112}\right|^2=\frac{8}{5}g_5^2$, implies  
$\left|g_{322}\right|^2=\left|g_{232}\right|^2=0$, which in turn leads to the following textures:
\begin{equation}
    \resizebox{0.92\textwidth}{!}{$M_u=\left(\begin{matrix}g_{112}\left\langle\mathcal{H}_2^5\right\rangle&g_{121}\left\langle\mathcal{H}_1^5\right\rangle&0\\g_{211}\left\langle\mathcal{H}_1^5\right\rangle&0&0\\0&0&g_{333}\left\langle\mathcal{H}_3^5\right\rangle\\\end{matrix}\right) \ \ , \ \ M_d=\left(\begin{matrix}0&0&0\\{\bar{g}}_{211}\left\langle{\bar{\mathcal{H}}}_{15}\right\rangle&{\bar{g}}_{223}\left\langle{\bar{\mathcal{H}}}_{35}\right\rangle&{\bar{g}}_{232}\left\langle{\bar{\mathcal{H}}}_{25}\right\rangle\\0&{\bar{g}}_{322}\left\langle{\bar{\mathcal{H}}}_{25}\right\rangle&{\bar{g}}_{333}\left\langle{\bar{\mathcal{H}}}_{35}\right\rangle\\\end{matrix}\right)~.$}
\end{equation}
These are clearly not phenomenologically viable options, so they are not
further analysed.  On the other hand, taking 
$\left|g_{112}\right|^2=\frac{4}{5}g_5^2$, which implies 
 $\left|{\bar{g}}_{322}\right|=0$, leads to:
\begin{equation}
\resizebox{0.92\textwidth}{!}{$M_u=\left(\begin{matrix}g_{112}\left\langle\mathcal{H}_2^5\right\rangle&g_{121}\left\langle\mathcal{H}_1^5\right\rangle&0\\g_{211}\left\langle\mathcal{H}_1^5\right\rangle&0&g_{232}\left\langle\mathcal{H}_2^5\right\rangle\\0&g_{322}\left\langle\mathcal{H}_2^5\right\rangle&g_{333}\left\langle\mathcal{H}_3^5\right\rangle\\\end{matrix}\right) \ \ , \ \ M_d=\left(\begin{matrix}{\bar{g}}_{112}\left\langle{\bar{\mathcal{H}}}_{25}\right\rangle&{\bar{g}}_{121}\left\langle{\bar{\mathcal{H}}}_{15}\right\rangle&0\\{\bar{g}}_{211}\left\langle{\bar{\mathcal{H}}}_{15}\right\rangle&{\bar{g}}_{223}\left\langle{\bar{\mathcal{H}}}_{35}\right\rangle&{\bar{g}}_{232}\left\langle{\bar{\mathcal{H}}}_{25}\right\rangle\\0&0&{\bar{g}}_{333}\left\langle{\bar{\mathcal{H}}}_{35}\right\rangle\\\end{matrix}\right) ~.$}\label{Modelo2SegundoCaso}
\end{equation}
%Although $M_u$ exhibits a viable texture, the form of $M_d$ does not
%look promising. 
In this case, $M_u$  exhibits a texture which is viable, but it is not as clear for $M_d$ \cite{Sharma:2015gfa,Xing:2020ijf}.  

Clearly, not taking these  particular extreme values recovers the general form of $V_3^{(2)}$, but with the novelty that the combination of the the two-loop finiteness conditions (\ref{ParametricSolModel2}) and the constraints imposed on the free parameters from eqs.~(\ref{eq:condV32}), allows for enough freedom to look for viable solutions. The sum rule will restrict the soft breaking terms, but also gives  flexibility when looking for a viable parameter space at low energies \cite{Heinemeyer:2019vbc}. Note that  the running of the RGE's will  certainly have an impact in the final results, and it can make a seemingly unviable texture at high energies viable at low energies and viceversa, in particular for large $\tan\beta$ \cite{Cakir:2006ut,Xing:2007fb}, which is our case.  The full phenomenological analysis will be left for a further publication.

\subsection{Model 3: Finite \texorpdfstring{$\mathbf{{V}_{4}^{\left({1}\right)}}$}{TEXT} at 2-loops}

In this case, we start from a texture of the form
$V_4^{\left(1\right)}$  for both mass matrices. To facilitate the
search for finite models, we  look for solutions with symmetries
that lead to zero texture mass matrices.  Such solutions
must lead to a hierarchical structure, with one of the Higgs
maximally coupled to the third generation. Naturally, there might be
other finiteness solutions to these type of mass matrices. The cyclic symmetries associated to
this model are shown in Table \ref{TablaModelo3}.

\begin{table}[H]
	\begin{center}
		\begin{tabular}{|c|c|c|c|c|c|c|c|c|c|c|c|c|c|c|c|}
			\hline
			& & & & & & & & & & & & & & & \\
			$Z_n$ & ${\bar{\Psi}}_1$ & ${\bar{\Psi}}_2$ & ${\bar{\Psi}}_3$ & $X_1$ & $X_2$ & $X_3$ & $\mathcal{H}_1$ & $\mathcal{H}_2$ & $\mathcal{H}_3$ & $\mathcal{H}_4$ & ${\bar{\mathcal{H}}}_1$ & ${\bar{\mathcal{H}}}_2$ & ${\bar{\mathcal{H}}}_3$ & ${\bar{\mathcal{H}}}_4$ & $\Sigma$ \\
			\hline
			$Z_2$ & $1$ & $1$ & $1$ & $1$ & $1$ & $1$ & $0$ & $0$ & $0$ & $0$ & $0$ & $0$ & $0$ & $0$ & $0$ \\
			\hline
			$Z_{8}$ & $4$ & $3$ & $5$ & $0$ & $7$ & $1$ & $0$ & $2$ & $6$ & $1$ & $4$ & $6$ & $2$ & $5$ & $0$ \\
			\hline
		\end{tabular}\caption{\label{TablaModelo3} Symmetries
                  related to (\ref{SolParametricModel3}).}
	\end{center}
      \end{table}
      
\noindent The following parametric solutions to (\ref{SistemaEcuacionesGeneral})
are found for this model:

\begin{displaymath}
\left|g_{124}\right|^2=\left|g_{214}\right|^2=\frac{4}{5}g_5^2\ \ ,\ \ \left|g_{222}\right|^2=\frac{2}{5}g_5^2\ \ ,\ \ \left|g_{231}\right|^2=\left|g_{321}\right|^2=\frac{1}{10}\left(8g_5^2-5\left|g_{111}\right|^2\right)\ \ ,\ \ 
\end{displaymath}
\begin{displaymath}
\left|g_{333}\right|^2=\frac{6}{5}g_5^2\ \ ,\ \ \left|{\bar{g}}_{111}\right|^2=\left|{\bar{g}}_{124}\right|^2=\frac{3}{20}\left(8g_5^2-5\left|g_{111}\right|^2\right)\ \ ,\ \ 
\end{displaymath}
\begin{displaymath}
\left|{\bar{g}}_{214}\right|^2=\frac{3}{4}\left|g_{111}\right|^2\ \ ,\ \ \left|{\bar{g}}_{222}\right|^2=\left|{\bar{g}}_{231}\right|^2=\frac{3}{10}g_5^2\ \ ,\ \ \left|{\bar{g}}_{321}\right|^2=-\frac{3}{20}\left(2g_5^2-5\left|g_{111}\right|^2\right)\ \ ,
\end{displaymath}
\begin{displaymath}
\ \ \left|{\bar{g}}_{333}\right|^2=\frac{9}{10}g_5^2\ \ ,\ \ \left|f_{22}\right|^2=\frac{3}{4}g_5^2\ \ ,\ \ \left|f_{33}\right|^2=\frac{g_5^2}{4}\ \ ,\ \ \left|p\right|^2=\frac{15}{7}g_5^2 \ \ ,
\end{displaymath}
\begin{equation}
\left|g_{132}\right|^2=\left|g_{312}\right|^2=\left|{\bar{g}}_{132}\right|^2=\left|{\bar{g}}_{312}\right|^2=\left|f_{11}\right|^2=\left|f_{44}\right|^2=0\ \ . 
\label{SolParametricModel3}
\end{equation}
By imposing the positivity conditon to the squared norm of the
couplings, we find the following constraint for $\left|g_{111}\right|^2$:
\begin{equation}
\frac{2}{5}g_5^2\le\left|g_{111}\right|^2\le\frac{8}{5}g_5^2 ~.
\label{IntervaloModelo3g111}
\end{equation}
Solutions with extra zero textures are found in the
limiting values for the coupling $\left|g_{111}\right|^2$. For
instance, when  $\left|g_{111}\right|^2=\frac{8}{5}g_5^2$, the
couplings $g_{321}$, $g_{231}$, ${\bar{g}}_{111}$ and
${\bar{g}}_{124}$ are zero, leading to the following mass textures:
\begin{equation}
\resizebox{0.92\textwidth}{!}{$
M_u=\left(\begin{matrix}g_{111}\left\langle\mathcal{H}_1^5\right\rangle&g_{124}\left\langle\mathcal{H}_4^5\right\rangle&0\\g_{214}\left\langle\mathcal{H}_4^5\right\rangle&g_{222}\left\langle\mathcal{H}_2^5\right\rangle&0\\0&0&g_{333}\left\langle\mathcal{H}_3^5\right\rangle\\\end{matrix}\right)\ \ ,\ \ M_d=\left(\begin{matrix}0&0&0\\{\bar{g}}_{214}\left\langle{\bar{\mathcal{H}}}_{45}\right\rangle&{\bar{g}}_{222}\left\langle{\bar{\mathcal{H}}}_{25}\right\rangle&{\bar{g}}_{231}\left\langle{\bar{\mathcal{H}}}_{15}\right\rangle\\0&{\bar{g}}_{321}\left\langle{\bar{\mathcal{H}}}_{15}\right\rangle&{\bar{g}}_{333}\left\langle{\bar{\mathcal{H}}}_{35}\right\rangle\\\end{matrix}\right) ~,$}
\end{equation}
clearly incompatible with phenomenology.
Another  solution is found when
$\left|{\bar{g}}_{321}\right|^2=0$, which leads to
$\left|g_{111}\right|^2=\frac{2}{5}g_5^2$  and to the following mass matrices:
\begin{equation}
\resizebox{0.92\textwidth}{!}{$
M_u=\left(\begin{matrix}g_{111}\left\langle\mathcal{H}_1^5\right\rangle&g_{124}\left\langle\mathcal{H}_4^5\right\rangle&0\\g_{214}\left\langle\mathcal{H}_4^5\right\rangle&g_{222}\left\langle\mathcal{H}_2^5\right\rangle&g_{231}\left\langle\mathcal{H}_1^5\right\rangle\\0&g_{321}\left\langle\mathcal{H}_1^5\right\rangle&g_{333}\left\langle\mathcal{H}_3^5\right\rangle\\\end{matrix}\right)\ \ ,\ \ M_d=\left(\begin{matrix}{\bar{g}}_{111}\left\langle{\bar{\mathcal{H}}}_{15}\right\rangle&{\bar{g}}_{124}\left\langle{\bar{\mathcal{H}}}_{45}\right\rangle&0\\{\bar{g}}_{214}\left\langle{\bar{\mathcal{H}}}_{45}\right\rangle&{\bar{g}}_{222}\left\langle{\bar{\mathcal{H}}}_{25}\right\rangle&{\bar{g}}_{231}\left\langle{\bar{\mathcal{H}}}_{15}\right\rangle\\0&0&{\bar{g}}_{333}\left\langle{\bar{\mathcal{H}}}_{35}\right\rangle\\\end{matrix}\right) ~,$}
\label{eq:mdmod3}
\end{equation}
which leads to a similar texture for $M_d$ as in
(\ref{Modelo2SegundoCaso}) of model 
$2$. Notice that since in the $M_d$ matrix of (\ref{eq:mdmod3}) the Higgs vev $\left\langle{\bar{\mathcal{H}}}_{35}\right\rangle$ only appears once, it is easier to accomodate for one heavy and two light down quarks, as compared to model 2.  

In the next section we explore another solution which
gives viable textures for both mass matrices. 

\subsection{Model 4: Finite \texorpdfstring{$\mathbf{{V}_{4}^{\left({2}\right)}}$}{TEXT}
  at 2-loops \label{model4}}

This model is defined by matrices of the type $V_4^{\left(2\right)}$:

\begin{equation}
\resizebox{0.92\textwidth}{!}{$M_u=\left(\begin{matrix}g_{114}\left\langle\mathcal{H}_4^5\right\rangle&g_{121}\left\langle\mathcal{H}_1^5\right\rangle&0\\g_{211}\left\langle\mathcal{H}_1^5\right\rangle&g_{224}\left\langle\mathcal{H}_4^5\right\rangle&g_{232}\left\langle\mathcal{H}_2^5\right\rangle\\0&g_{322}\left\langle\mathcal{H}_2^5\right\rangle&g_{333}\left\langle\mathcal{H}_3^5\right\rangle\\\end{matrix}\right)\ ,\ \ M_d=\left(\begin{matrix}{\bar{g}}_{114}\left\langle{\bar{\mathcal{H}}}_{45}\right\rangle&{\bar{g}}_{121}\left\langle{\bar{\mathcal{H}}}_{15}\right\rangle&0\\{\bar{g}}_{211}\left\langle{\bar{\mathcal{H}}}_{15}\right\rangle&{\bar{g}}_{224}\left\langle{\bar{\mathcal{H}}}_{45}\right\rangle&{\bar{g}}_{232}\left\langle{\bar{\mathcal{H}}}_{25}\right\rangle\\0&{\bar{g}}_{322}\left\langle{\bar{\mathcal{H}}}_{25}\right\rangle&{\bar{g}}_{333}\left\langle{\bar{\mathcal{H}}}_{35}\right\rangle\\\end{matrix}\right)\ ~.$}
	\label{MatricesMasaModelo4}
\end{equation}
These matrices have the advantage that there is only one Higgs
corresponding to $\left(\mathcal{H}_3,{\bar{\mathcal{H}}}_3 \right)$
in the entry $\left(M_f\right)_{33}$, with $f=u,d$.  This helps build a model with a heavy quark in the third family, while leaving
the first two families light. The matrices already have one texture zero,
which helps to find non-parametric solutions to the finiteness
conditions.
Thus, we have the following finiteness solutions to  
(\ref{SistemaEcuacionesGeneral}):
\begin{displaymath}
\left|g_{211}\right|^2=\left|g_{121}\right|^2 \ \ , \ \ \left|g_{232}\right|^2=\frac{1}{5}\left(16g_5^2-5\left|g_{114}\right|^2-10\left|g_{121}\right|^2-5\left|g_{224}\right|^2\right) \ , 
\end{displaymath}
\begin{displaymath}
\left|g_{312}\right|^2=\left|g_{132}\right|^2=0 \ \ , \ \ \left|g_{322}\right|^2=\frac{1}{5}\left(16g_5^2-5\left|g_{114}\right|^2-10\left|g_{121}\right|^2-5\left|g_{224}\right|^2\right) \ ,
\end{displaymath}
\begin{displaymath}
\left|g_{333}\right|^2=\frac{1}{5}(-8g_5^2+5\left|g_{114}\right|^2+10\left|g_{121}\right|^2+5\left|g_{224}\right|^2) \ ,
\end{displaymath}
\begin{displaymath}
\left|{\bar{g}}_{114}\right|^2=\frac{3}{20}\left(16g_5^2-5\left|g_{114}\right|^2-10\left|g_{121}\right|^2\right) \ \ , \ \ \left|{\bar{g}}_{121}\right|^2=\frac{3}{20}\left(8g_5^2-5\left|g_{114}\right|^2\right) \ , 
\end{displaymath}
\begin{displaymath}
\left|{\bar{g}}_{211}\right|^2=-\frac{3}{20}\left(8g_5^2-5\left|g_{114}\right|^2-10\left|g_{121}\right|^2\right) \ , 
\end{displaymath}
\begin{displaymath}
\left|{\bar{g}}_{224}\right|^2=-\frac{3}{20}\left(16g_5^2-10\left|g_{114}\right|^2-10\left|g_{121}\right|^2-5\left|g_{224}\right|^2\right) \ ,
\end{displaymath}
\begin{displaymath}
\left|{\bar{g}}_{224}\right|^2=-\frac{3}{20}\left(16g_5^2-10\left|g_{114}\right|^2-10\left|g_{121}\right|^2-5\left|g_{224}\right|^2\right) \ ,    
\end{displaymath}
\begin{displaymath}
\left|{\bar{g}}_{232}\right|^2=\frac{3}{20}\left(16g_5^2-5\left|g_{114}\right|^2-10\left|g_{121}\right|^2-5\left|g_{224}\right|^2\right) \ ,
\end{displaymath}
\begin{displaymath}
\left|{\bar{g}}_{322}\right|^2=\frac{3}{20}\left(16g_5^2-5\left|g_{114}\right|^2-10\left|g_{121}\right|^2-5\left|g_{224}\right|^2\right) \ ,
\end{displaymath}
\begin{displaymath}
\left|{\bar{g}}_{333}\right|^2=-\frac{3}{20}\left(8g_5^2-5\left|g_{114}\right|^2-10\left|g_{121}\right|^2-5\left|g_{224}\right|^2\right) \ ,
\end{displaymath}
\begin{displaymath}
	\left|f_{11}\right|^2=\frac{1}{4}\left(4g_5^2-5\left|g_{121}\right|^2\right) \ \ , \ \ \left|f_{22}\right|^2=\frac{1}{4}\left(-12g_5^2+5\left|g_{114}\right|^2+10\left|g_{121}\right|^2+5\left|g_{224}\right|^2\right) \ ,
\end{displaymath}
\begin{displaymath}
	\left|f_{33}\right|^2=\frac{1}{8}\left(16g_5^2-5\left|g_{114}\right|^2-10\left|g_{121}\right|^2-5\left|g_{224}\right|^2\right) \ ,
\end{displaymath}
\begin{equation}
\left|f_{44}\right|^2=\frac{1}{8}\left(8g_5^2-5\left|g_{114}\right|^2-5\left|g_{224}\right|^2\right) \ \ , \ \ \left|p\right|^2=\frac{15}{7}g_5^2 \ .
 \label{SolParametricModel4}
\end{equation}
Although the form of the mass matrices suggests searching for Nearest Neighbour Interaction (NNI) type
matrices, the finiteness conditions lead to solutions with negative
squared norms. On the other hand, it is possible to find solutions
where  $\left(M_f\right)_{22}=0$ and that also have
$\left(M_f\right)_{11}$ small. When $\left|g_{224}\right|^2=0$ and 
$\left|{\bar{g}}_{224}\right|^2=0$, the parametric solutions simplify
to:  

\begin{displaymath}
\left|g_{121}\right|^2=\frac{1}{5}\left(8g_5^2-5\left|g_{114}\right|^2\right) \ \ , \ \ \left|g_{211}\right|^2=\frac{1}{5}\left(8g_5^2-5\left|g_{114}\right|^2\right) \ ,
\end{displaymath}
\begin{displaymath}
\left|g_{224}\right|^2=0 \ \ , \ \ \left|g_{232}\right|^2=\left|g_{114}\right|^2 \ \ , \ \ \left|g_{312}\right|^2=\left|g_{132}\right|^2=0 \ ,
\end{displaymath}
\begin{displaymath}
\left|g_{322}\right|^2=\left|g_{114}\right|^2 \ \ , \ \ \left|g_{333}\right|^2=\frac{1}{5}\left(8g_5^2-5\left|g_{114}\right|^2\right) \ ,
\end{displaymath}
\begin{displaymath}
\left|{\bar{g}}_{114}\right|^2=\frac{3\left|g_{114}\right|^2}{4} \ \ , \ \ \left|{\bar{g}}_{121}\right|^2=\frac{3}{20}\left(8g_5^2-5\left|g_{114}\right|^2\right) \ ,
\end{displaymath}
\begin{displaymath}
\left|{\bar{g}}_{211}\right|^2=\frac{3}{20}\left(8g_5^2-5\left|g_{114}\right|^2\right) \ \ , \ \ \left|{\bar{g}}_{224}\right|^2=0 \ \ , \ \ \left|{\bar{g}}_{232}\right|^2=\frac{3\left|g_{114}\right|^2}{4} \ ,
\end{displaymath}
\begin{displaymath}
\left|{\bar{g}}_{322}\right|^2=\frac{3\left|g_{114}\right|^2}{4} \ \ , \ \ \left|{\bar{g}}_{333}\right|^2=\frac{3}{20}\left(8g_5^2-5\left|g_{114}\right|^2\right) \ ,
\end{displaymath}
\begin{displaymath}
\left|f_{11}\right|^2=\frac{1}{4}\left(-4g_5^2+5\left|g_{114}\right|^2\right) \ \ , \ \ \left|f_{22}\right|^2=\frac{1}{4}\left(4g_5^2-5\left|g_{114}\right|^2\right) \ \ , \ \ \left|f_{33}\right|^2=\frac{5\left|g_{114}\right|^2}{8}
\end{displaymath}
\begin{equation}
\left|f_{44}\right|^2=\frac{1}{8}\left(8g_5^2-5\left|g_{114}\right|^2\right) \ \ , \ \ \left|p\right|^2=\frac{15}{7}g_5^2 ~.
\label{Sol2ParametricModel4}
\end{equation}
that leads to $\left|f_{11}\right|^2=\left|f_{22}\right|^2=0$ for
consistency.  This in turn leads to the all-loop finite model 4.1,
presented in next section.

The mass matrices 
(\ref{MatricesMasaModelo4}) could have up to 6 phases, in the entries 
 $(M_{u})_{11}$, $(M_{u})_{22}$,
$(M_{u})_{33}$, $(M_{d})_{21}$, $(M_{d})_{32}$ and  $(M_{d})_{33}$.
But in the solutions (\ref{Sol2ParametricModel4}), only four phases are
necessary, placed in the position of the original textures 
(\ref{MatricesMasaModelo4}). The resulting textures, presented in next
section,  have the potential
to reproduce the quark masses and mixing.

\subsection{Model 4.1: Finite version at all orders of \texorpdfstring{$\mathbf{{V}_{4}^{\left({2}\right)}}$}{TEXT}}

The following model is similar to the one presented in ref. \cite{Babu:2002in}, but in our case it exhibits only cyclic symmetries.  Another difference lies in the inclusion of the phases and their position, not previously done in FUTs.  We found other models also based in cyclic symmetries and finite to all-loops with different textures to model 4.1, but it can be shown that they all belong to the same equivalence class, so they are basically the same model.

To ensure finiteness to all-loops, the solutions to 
(\ref{Sol2ParametricModel4})  must be isolated and non-degenerate. One
such solution exhibits the symmetries presented in Table
\ref{TablaSimetriasCiclicasModelo4.1}, where it should be noted that they lead
to $\left|g_{224}\right|^2=\left|{\bar{g}}_{224}\right|^2=\left|f_{11}\right|^2=0$.

\begin{table}[t]
	\begin{center}
		\resizebox{14cm}{!}{
			\begin{tabular}{|c|c|c|c|c|c|c|c|c|c|c|c|c|c|c|c|}
				\hline
				& & & & & & & & & & & & & & &
				\\
				$Z_{n}$ & $\bar{\Psi}_{1}$ & $\bar{\Psi}_{2}$ & $\bar{\Psi}_{3}$ & $X_{1}$ & $X_{2}$ & $X_{3}$ & $\mathcal{H}_{1}$ & $\mathcal{H}_{2}$ & $\mathcal{H}_{3}$ & $\mathcal{H}_{4}$ & $\mathcal{\bar{H}}_{1}$ & $\mathcal{\bar{H}}_{2}$ & $\mathcal{\bar{H}}_{3}$ & $\mathcal{\bar{H}}_{4}$ & $\Sigma$ \\
				\hline
				$Z_{2}$ & $1$ & $1$ & $1$ & $1$ & $1$ & $1$ & $0$ & $0$ & $0$ & $0$ & $0$ & $0$ & $0$ & $0$ & $0$ \\
				\hline
				$Z_{3}$ & $0$ & $2$ & $0$ & $0$ & $2$ & $0$ & $1$ & $1$ & $0$ & $0$ & $1$ & $1$ & $0$ & $0$ & $0$ \\
				\hline
				$Z_{4}$ & $3$ & $3$ & $2$ & $3$ & $3$ & $2$ & $2$ & $3$ & $0$ & $2$ & $2$ & $3$ & $0$ & $2$ & $0$ \\
				\hline
			\end{tabular}
		}
		{	\caption{\label{TablaSimetriasCiclicasModelo4.1} Cyclic discrete symmetries of model $4.1$ to obtain isolated, non-degenerate solutions of the system of equations (\ref{SistemaEcuacionesGeneral}).
		}}
	\end{center}
\end{table}	
\noindent Under these symmetries, the superpotential is:
\begin{dmath}
\mathcal{W}_{\textrm{4.1}}={\bar{g}}_{114}{\bar{\Psi}}_{1i}X_1^{ij}{\bar{\mathcal{H}}}_{4j}+{\bar{g}}_{121}{\bar{\Psi}}_{2i}X_1^{ij}{\bar{\mathcal{H}}}_{1j}+{\bar{g}}_{211}{\bar{\Psi}}_{1i}X_2^{ij}{\bar{\mathcal{H}}}_{1j}+{\bar{g}}_{232}{\bar{\Psi}}_{3i}X_2^{ij}{\bar{\mathcal{H}}}_{2j}+{\bar{g}}_{322}{\bar{\Psi}}_{2i}X_3^{ij}{\bar{\mathcal{H}}}_{2j}+{\bar{g}}_{333}{\bar{\Psi}}_{3i}X_3^{ij}{\bar{\mathcal{H}}}_{3j}+\frac{1}{2}g_{114}\epsilon_{ijklm}X_1^{ij}X_1^{kl}\mathcal{H}_4^m+g_{121}\epsilon_{ijklm}X_1^{ij}X_2^{kl}\mathcal{H}_1^m+g_{232}\epsilon_{ijklm}X_2^{ij}X_3^{kl}\mathcal{H}_2^m+\frac{1}{2}g_{333}\epsilon_{ijklm}X_3^{ij}X_3^{kl}\mathcal{H}_3^m+f_{33}{\bar{\mathcal{H}}}_{3j}\Sigma_{\ i}^j\mathcal{H}_3^i+f_{44}{\bar{\mathcal{H}}}_{4j}\Sigma_{\ i}^j\mathcal{H}_4^i+\frac{1}{3!}p\Sigma_{\ j}^i\ \Sigma_{\ k}^j\ \Sigma_{\ i}^k+\frac{1}{2}\lambda^{\left(\Sigma\right)}\Sigma_{\ j}^i\Sigma_{\ i}^j+m_{ab}{\bar{\mathcal{H}}}_{ai}\mathcal{H}_b^i \textrm{   .}
	\label{SuperpotencialModelo4.1}
\end{dmath}
The trilinear terms satisfy exactly these symmetries, but the term 
$m_{ab}{\bar{\mathcal{H}}}_{ai}\mathcal{H}_b^i$ breaks them softly in
order to ensure a successful doublet-triplet splitting and no
fast proton decay.  This breaking does not affect the finiteness
conditions, which apply only to the trilinear terms in the superpotential.

\noindent Using the allowed couplings in  (\ref{SuperpotencialModelo4.1}), we
obtain the following solutions:
\begin{displaymath}
\left|g_{114}\right|^2=\left|g_{121}\right|^2=\left|g_{211}\right|^2=\left|g_{232}\right|^2=\left|g_{322}\right|^2=\left|g_{333}\right|^2=\frac{4}{5}g_5^2  \ \ \ \textrm{   ,}
\end{displaymath}
\begin{displaymath}
	\left|{\bar{g}}_{114}\right|^2=\left|{\bar{g}}_{121}\right|^2=\left|{\bar{g}}_{211}\right|^2=\left|{\bar{g}}_{232}\right|^2=\left|{\bar{g}}_{322}\right|^2=\left|{\bar{g}}_{333}\right|^2=\frac{3}{5}g_5^2  \ \ \ \textrm{   ,}
\end{displaymath}
\begin{equation}
	\left|f_{33}\right|^2=\left|f_{44}\right|^2=\frac{1}{2}g_5^2 \ \ \  \textrm{   ,} \ \ \	\left|p\right|^2=\frac{15}{7}g_5^2 \ \ \  \textrm{   .}
	\label{SolFinitudModelo4.1}
\end{equation}
Since these solutions are unique, isolated and non-degenerate, the
model is all-loop finite.

\subsubsection{Sum rule for squared masses of scalars}

In the dimesionful sector the one-loop sum rule  (\ref{1loopsumrule}) generates
13 equations, with 16 parameters $(m_{{\widetilde{\bar{\psi}}}_1}^2$,
$m_{{\widetilde{\bar{\psi}}}_2}^2$,
$m_{{\widetilde{\bar{\psi}}}_3}^2$, $m_{{\widetilde{\chi}}_1}^2$,
$m_{{\widetilde{\chi}}_2}^2$, $m_{{\widetilde{\chi}}_3}^2$,
$m_{H_1}^2$, $m_{H_2}^2, m_{H_3}^2$, $m_{H_4}^2$, $m_{{\bar{H}}_1}^2$,
$m_{{\bar{H}}_2}^2$, $m_{{\bar{H}}_3}^2$, $m_{{\bar{H}}_4}^2$,
$m_{\phi_\Sigma}^2$ y $M$), which in this model lead to only three
independent parameters 
($m_{H_1}^2$,  $m_{H_3}^2$, $M$), as shown in
(\ref{SolucionesReglaSumaModelo}). The sum rules in terms of these three
parameters are:
\begin{displaymath}
m_{{\widetilde{\bar{\psi}}}_1}^2=m_{{\widetilde{\bar{\psi}}}_3}^2=\frac{1}{6}\left(-MM^\dag+9m_{H_3}^2\right)\ \ \ ,\ \ \ m_{{\widetilde{\bar{\psi}}}_2}^2=\frac{1}{6}\left(-MM^\dag-6m_{H_1}^2+15m_{H_3}^2\right)\ ,
\end{displaymath}
\begin{displaymath}
m_{{\widetilde{\chi}}_1}^2=m_{{\widetilde{\chi}}_3}^2=\frac{1}{2}\left(MM^\dag-m_{H_3}^2\right)\ \ \ ,\ \ \ m_{{\widetilde{\chi}}_2}^2=\frac{1}{2}\left(MM^\dag-2m_{H_1}^2+m_{H_3}^2\right)\ ,
\end{displaymath}
\begin{displaymath}
	m_{{\bar{H}}_1}^2=m_{{\bar{H}}_2}^2=\frac{1}{3}\left(2MM^\dag+3m_{H_1}^2-6m_{H_3}^2\right)\ \ \ ,\ \ \ m_{{\bar{H}}_3}^2=m_{{\bar{H}}_4}^2=\frac{1}{3}\left(2MM^\dag-3m_{H_3}^2\right)\ ,
\end{displaymath}
\begin{equation}
m_{H_2}^2=m_{H_1}^2\ \ \ ;\ \ \ m_{H_4}^2=m_{H_3}^2\ \ \ ,\ \ \ m_{\phi_\Sigma}^2=\frac{1}{3}MM^\dag\ \ \ . \ \ \
\label{SolucionesReglaSumaModelo}
\end{equation}

\subsubsection{Invariant products under phase transformations}

Since the finiteness conditions only restrict the squared norm of the
couplings,  we address the possible phases in the
least arbitrary possible way.  The minimum amount of phases needed and
their positions in the mass matrices have to satisfy certain theorems
and requirements established in references
\cite{Kusenko:1993wu,Kusenko:1994zm}, which we follow here.

For a three generation quark model, there exist several combinations
of four 
($P_{4;j_{1}k_{1},j_{2}k_{2}}^{\left(f\right)}$,
$Q_{4;j_{1}k_{1},j_{2}m_{1}}$) or even six 
($P_{6;j_1k_1,j_2k_2,j_3k_3}^{\left(f\right)}$,
$Q_{6;j_1k_1,j_2k_2,j_3m_1}^{\left(fff^\prime\right)}$) entries of the
Yukawa matrices that are needed to determine the position of the
phases. Their argument must be different from zero and 
$\pi$, and remains invariant under rephasing of the Yukawa
couplings. The general form of these products is the following: 
\begin{displaymath}
P_{4;j_1k_1,j_2k_2}^{\left(f\right)}=Y_{f,j_1k_1}Y_{f,j_2k_2}Y_{f,j_2k_1}^\ast Y_{f,j_1k_2}^\ast\ \ \ \ ,\ \ \ \ Q_{4;j_1k_1,j_2m_1}=Y_{u,j_1k_1}Y_{d,j_2m_1}Y_{u,j_2k_1}^\ast Y_{d,j_1m_1}^\ast
\end{displaymath}
\begin{displaymath}
P_{6;j_1k_1,j_2k_2,j_3k_3}^{\left(f\right)}=Y_{f,j_1k_1}Y_{f,j_2k_2}Y_{f,j_3k_3}Y_{f,j_2k_1}^\ast Y_{f,j_3k_2}^\ast Y_{f,j_1k_3}^\ast\ \ \ ,
\end{displaymath}
\begin{equation}
Q_{6;j_1k_1,j_2k_2,j_3m_1}^{\left(fff^\prime\right)}=Y_{f,j_1k_1}Y_{f,j_2k_2}Y_{f^\prime,j_3m_1}Y_{f,j_1k_2}^\ast Y_{f,j_3k_1}^\ast Y_{f^\prime,j_2m_1}^\ast\ \ \ ,
\end{equation}
where $\left(fff^\prime\right)=\left(uud\right)$ or
$\left(fff^\prime\right)=\left(ddu\right)$.
The argument of the above products is equivalent to a system of
coupled equations among the phases. Thus, once the minimum number of
phases is established it is enough to consider the same number of
products that generate a system of linearly independent equations.

For this model, the minimum number of phases is 4 among both types of Yukawa
matrices and their positions are established by constructing the
following invariant products under phase transformations and
extracting their arguments:
\begin{displaymath}
\arg{\left(g_{114}{\bar{g}}_{211}g_{211}^\ast{\bar{g}}_{114}^\ast\right)}=C_1\neq0,\pi\ \ \ ,\ \ \ \arg{\left(g_{121}{\bar{g}}_{322}g_{322}^\ast{\bar{g}}_{121}^\ast\right)}=C_2\neq0,\pi  ~~,
\end{displaymath}
\begin{equation}
\arg{\left(g_{232}{\bar{g}}_{333}g_{333}^\ast{\bar{g}}_{232}^\ast\right)}=C_3\neq0,\pi\ \ \ ,\ \ \ \arg{\left(g_{114}g_{232}g_{322}g_{121}^\ast g_{211}^\ast g_{333}^\ast\right)}=C_4\neq0,\pi\ \ .
\label{ProductosInvariantesModelo4.1}
\end{equation}
These expressions are dependent on 12 Yukawa couplings, but since
$Y_u$ is symmetric there are only 10 left, of which 6 are real.
Therefore we are left with the following solutions for the phases:
\begin{displaymath}
\arg{\left(g_{333}\right)}=-C_4+\arg{\left(g_{114}\right)}-\arg{\left(g_{121}\right)}-\arg{\left(g_{211}\right)}+\arg{\left(g_{232}\right)}+\arg{\left(g_{322}\right)}=\phi_3 ~,
\end{displaymath}
\begin{displaymath}
\arg{\left({\bar{g}}_{211}\right)}=C_1-\arg{\left(g_{114}\right)}+\arg{\left(g_{211}\right)}+\arg{\left({\bar{g}}_{114}\right)}={\bar{\phi}}_1 ~,
\end{displaymath}
\begin{displaymath}
\arg{\left({\bar{g}}_{322}\right)}=C_2-\arg{\left(g_{121}\right)}+\arg{\left(g_{322}\right)}+\arg{\left({\bar{g}}_{121}\right)}={\bar{\phi}}_2 ~,
\end{displaymath}
\begin{equation}
\arg{\left({\bar{g}}_{333}\right)}=C_3-C_4+\arg{\left(g_{114}\right)}-\arg{\left(g_{121}\right)}-\arg{\left(g_{211}\right)}+\arg{\left(g_{322}\right)}+\arg{\left({\bar{g}}_{232}\right)}={\bar{\phi}}_3 ~.
\label{PosicionFasesModelo4.1}
\end{equation}
This way we choose  $\phi_3=\arg{\left(g_{333}\right)}$,
${\bar{\phi}}_1=\arg{\left({\bar{g}}_{211}\right)}$,
${\bar{\phi}}_2=\arg{\left({\bar{g}}_{322}\right)}$ y
${\bar{\phi}}_3=\arg{\left({\bar{g}}_{333}\right)}$
to fix the phases in the mass matrices.  There are various ways to
distribute the four phases, but in this analysis we choose this
particular one.

\subsubsection{Mass matrices}

The mass matrices for this model are:
\begin{equation}
	\resizebox{0.92\textwidth}{!}{$M_u=\left(\begin{matrix}g_{114}\left\langle\mathcal{H}_4^5\right\rangle&g_{121}\left\langle\mathcal{H}_1^5\right\rangle&0\\g_{211}\left\langle\mathcal{H}_1^5\right\rangle&0&g_{232}\left\langle\mathcal{H}_2^5\right\rangle\\0&g_{322}\left\langle\mathcal{H}_2^5\right\rangle&g_{333}\left\langle\mathcal{H}_3^5\right\rangle\\\end{matrix}\right)=\frac{2}{\sqrt5}g_5\left(\begin{matrix}\left\langle\mathcal{H}_4^5\right\rangle&\left\langle\mathcal{H}_1^5\right\rangle&0\\\left\langle\mathcal{H}_1^5\right\rangle&0&\left\langle\mathcal{H}_2^5\right\rangle\\0&\left\langle\mathcal{H}_2^5\right\rangle&e^{i\phi_3}\left\langle\mathcal{H}_3^5\right\rangle\\\end{matrix}\right)\
	\textrm{   ,}$}
\end{equation}

\begin{equation}
\resizebox{0.92\textwidth}{!}{$M_d=\left(\begin{matrix}{\bar{g}}_{114}\left\langle{\bar{\mathcal{H}}}_{45}\right\rangle&{\bar{g}}_{121}\left\langle{\bar{\mathcal{H}}}_{15}\right\rangle&0\\{\bar{g}}_{211}\left\langle{\bar{\mathcal{H}}}_{15}\right\rangle&0&{\bar{g}}_{232}\left\langle{\bar{\mathcal{H}}}_{25}\right\rangle\\0&{\bar{g}}_{322}\left\langle{\bar{\mathcal{H}}}_{25}\right\rangle&{\bar{g}}_{333}\left\langle{\bar{\mathcal{H}}}_{35}\right\rangle\\\end{matrix}\right)=\sqrt{\frac{3}{5}}g_5\left(\begin{matrix}\left\langle{\bar{\mathcal{H}}}_{45}\right\rangle&\left\langle{\bar{\mathcal{H}}}_{15}\right\rangle&0\\e^{i{\bar{\phi}}_1}\left\langle{\bar{\mathcal{H}}}_{15}\right\rangle&0&\left\langle{\bar{\mathcal{H}}}_{25}\right\rangle\\0&e^{i{\bar{\phi}}_2}\left\langle{\bar{\mathcal{H}}}_{25}\right\rangle&e^{i{\bar{\phi}}_3}\left\langle{\bar{\mathcal{H}}}_{35}\right\rangle \\ \end{matrix}\right)
	\textrm{   .}$}
\end{equation}
We have already substituted in these matrices the solutions found for the
finiteness conditions (\ref{SolFinitudModelo4.1}) and the complex
phases, as already explained above.

\noindent After the rotation in the Higgs sector, the matrices in the MSSM basis
are:
\begin{equation}
M_u=\frac{2}{\sqrt5}g_5\left(\begin{matrix}{\widetilde{\alpha}}_4&{\widetilde{\alpha}}_1&0\\{\widetilde{\alpha}}_1&0&{\widetilde{\alpha}}_2\\0&{\widetilde{\alpha}}_2&e^{i\phi_3}{\widetilde{\alpha}}_3\\\end{matrix}\right)\left\langle\mathcal{K}_3^5\right\rangle\ ,
\end{equation}
\begin{equation}
M_d=\sqrt{\frac{3}{5}}g_5\left(\begin{matrix}{\widetilde{\beta}}_4&{\widetilde{\beta}}_1&0\\e^{i{\bar{\phi}}_1}{\widetilde{\beta}}_1&0&{\widetilde{\beta}}_2\\0&e^{i{\bar{\phi}}_2}{\widetilde{\beta}}_2&e^{i{\bar{\phi}}_3}{\widetilde{\beta}}_3\\\end{matrix}\right)\left\langle{\bar{\mathcal{K}}}_{35}\right\rangle\ ,
\end{equation}
where $\widetilde{\alpha}_i$ and $\widetilde{\beta}_i$ refer to the
rotation angles in the up and down sector, respectively.  

Notice that this is a non-minimal SUSY $SU(5)$ model. The actual expressions for the five dimensional operators that mediate proton decay  will depend, among other parameters, on the Yukawa couplings, the $V_{CKM}$ matrix elements, the soft breaking sector, and the coloured Higgs triplet masses, and will differ from the usual minimal $SU(5)$ \cite{Nath:2007eg,Babu:2013jba}.  It is possible to estimate though, that the coloured triplets are indeed heavier than the GUT scale in this model, similarly to the diagonal FUTA model.   In refs.\cite{Arnowitt:1993pd,Nath:2007eg} a way to suppress these five dimensional operators in models with several heavy triplets is outlined, in the basis where  only one pair of Higgs doublets couples to matter, similar to the scenario we present here. 

\subsubsection{Free parameters \label{CantidadParametrosModelo4.1}}
Before the solution to the finiteness conditions is determined, the
Lagrangian has 89 free parameters, including all the couplings,
soft breaking terms, and phases, plus the vacuum expectation values of
the Higgs fields. This number is drastically reduced after the
solution to the finiteness conditons is found, both in the
dimensionless and dimensionful sectors, leaving 33 free parameters. This
number is further
reduced by imposing the doublet-triplet splitting, which again has
consequences in the dimensionless and soft breaking sectors. The four
phases that are left as free parameters are constrained by the
invariant products, as already explained, as:
\begin{equation}
{\bar{\phi}}_1\neq0,\pi\ \ \ ;\ \ \ {\bar{\phi}}_2\neq0,\pi\ \ \ ;\ \ \ {\bar{\phi}}_3\neq0,\pi\ \ \ ;\ \ \ \phi_3\neq0,\pi\ \ \ ;\ \ \ {\bar{\phi}}_3-\phi_3\neq0,\pi ~.
\end{equation}

Then, the vacuum
expectation values of the Higgs fields are replaced by the rotation
angles when we go to the MSSM basis, with the constraint that the 
sum of their squared values is equal to one, which eliminates one more parameter
\begin{equation}
{\widetilde{\alpha}}_4=\sqrt{1-{\widetilde{\alpha}}_1^2-{\widetilde{\alpha}}_2^2-{\widetilde{\alpha}}_3^2} \ \ \textrm{,} \ \  {\widetilde{\beta}}_4=\sqrt{1-{\widetilde{\beta}}_1^2-{\widetilde{\beta}}_2^2-{\widetilde{\beta}}_3^2} ~.
\end{equation}
Once we match our model with the
MSSM, there are further reductions in the number of free parameters by
requiring radiative electroweak symmetry breaking and by fixing the
value of the tau mass at low energies, which together with the value
of the SM vev, fixes $v_d$ and $\tan\beta$.  Finally, we are left with
12 parameters in total, including the soft breaking terms, the
phases, and the 
rotation angles:
\begin{equation}
{\widetilde{\alpha}}_1, \ {\widetilde{\alpha}}_2, \ {\widetilde{\alpha}}_3, \ {\widetilde{\beta}}_1, \ {\widetilde{\beta}}_2, \ {\widetilde{\beta}}_3, \ \phi_3, \ {\bar{\phi}}_1, \ {\bar{\phi}}_2, \ {\bar{\phi}}_3, \ M, \ \mu ~.
\end{equation}
It should be noted that only one combination of the phases is an observable, reducing further the number of free parameters, but we present the four of them here. The form of the $V_{CKM}$ matrix will single out this combination.

\section{Discussion and Conclusions} %Conclusions}

In the present work, we analyzed four $SU\left(5\right)$ FUT models with discrete flavour symmetries $S_3$ and $Z_N$, which are finite at two-loop or to all-loop orders. The first model has an $S_3\times Z_2\times Z_3$ flavour symmetry, where two generations of quarks are assigned to an $S_3$ doublet irrep, and the third to the singlet, whereas in the Higgs sector two Higgs pairs are assigned to a doublet, one pair to the symmetric singlet and one to the antisymmetric singlet of $S_3$.   The model is finite to all-orders, but leads to a phenomenologically non-viable mass matrix. Nevertheless, different requirements or constraints in the $S_3$ model, or a different non-Abelian group, could lead to mass matrices which are phenomenologically viable.

The other models exhibit  cyclic symmetries and were obtained from the analysis of eight different matrix structures \cite{Babu:2002in}, that guarantee that some of the off-diagonal elements are zero and where it was assumed that $M_u$ and $M_d$ would have the same texture. From this analysis, we chose three textures that looked promising to reproduce the quark masses and mixing pattern, models 2, 3, and 4.  In model 2  only three pairs of Higgs fields couple to the quarks, whereas in models 3 and 4 all four pairs of Higgs doublets couple to the quarks.  In the three models, parametric solutions to the finiteness conditions were found, which imply two-loop finiteness. There are two free parameters in case of model 2 and one in model 3, and a range of validity is found for these to ensure the positivity of the squared norm of the Yukawa couplings. Evaluating the free parameters in the allowed range gives a particular texture with no more free parameters,  except for the phases, which are in principle phenomenologically viable.
%Both models could lead to particular viable textures for the mass matrices. 
Taking the exact limiting values in each range  lead to mass matrices with more texture zeroes, but in these particular cases the solutions do not appear viable, although to fully establish this fact a running to low energies and the inclusion of the soft breaking terms with the sum rule should be performed.

Finally, with texture number 4 we found an all-loop solution, labeled 4.1. This model exhibits the same texture as the one found in ref. \cite{Babu:2002in}, although in our case it exhibits different discrete symmetries, which could affect the bilinear terms in the soft breaking sector, not analyzed here.  Another difference is that we do consider that the Yukawa couplings are complex, and assign the number and position of the phases in the mass matrices using invariant products under phase transformations, which allows us to reduce and constrain further the number of free parameters. In this model, all four pairs of Higgs doublets are coupled to the quarks, and, similarly to model 3, there is only one term in the $(3,3)$ entry of the mass matrices. This makes it more feasible to achieve one heavy and two light eigenvalues  in the mass matrices, thus making it the most phenomenologically promising model. Moreover, for this $(3,3)$ entry, corresponding to the third generation of quarks, the values of this model are the same ones as for model FUTB, which is known to give accurate predictions for the top and bottom masses \cite{Heinemeyer:2019vbc}. The proposed hierarchy for the mass matrix elements allows approximating the texture either to a diagonal one, a block diagonal one, or a NNI matrix.  It is worth noting that this model also has the greatest reduction in the number of free parameters. 

We only address here the flavour problem in the quark sector. The inclusion of neutrino masses in this type of models is not straightforward given that the matter content is restricted and adding a $U(1)$ gauge group is not allowed by the finitess conditions, but it may be done by introducing bilinear terms that break R parity (see for instance \cite{Hall:1983id,Hempfling:1995wj,Hirsch:2000ef,Choudhury:2023lbp}).  This would not affect directly the solutions to the finiteness conditions for the fermions, but may have an indirect effect through the soft supersymmetry breaking effects.  %In these scenarios it is possible to choose these R parity violation (RPV) terms to get the neutrino masses in the right region 
 On the other hand, it is also possible not only to generate the neutrino masses but also to correct the wrong mass relations between charged leptons and down type quarks that appear in $SU(5)$ via R parity violation (RPV) \cite{Bajc:2015zja}.  An alternative to fix this latter problem is to use the SUSY loop corrections in different ways to obtain the correct charged leptons and down type masses, as was done for instance in refs. \cite{Babu:1989fg,Bajc:2015ita,Diaz-Cruz:2000nvf,Ross:2007az}.  Also, the determination of the soft spectrum will have an impact on the solution to the proton decay problem inherent to $SU(5)$, and could be suppressed more naturally  in a type of split-SUSY scenario, as for example  \cite{Hisano:2022qll}. One other possibility to avoid proton decay is to set the mass of  the coloured Higgs triplets close to the Planck scale, although such a fine tuning would also have an impact on the tuning necessary in the soft breaking sector. All these  scenarios  could in principle be adapted to the models presented here.

A full phenomenological analysis of the most promising models will be left for a future publication. In this respect it is important to stress that for this analysis a proper implementation of the soft breaking sector using the sum rule has to be performed, and that after the RG running the apparent viability or unviability of a texture at high energies might change \cite{Cakir:2006ut,Xing:2007fb}. 

\section*{Acnowledgements}
LOER and MM acknowledge support from UNAM project PAPIIT  IN111224 and CONAHCYT project CBF2023-2024-548. {LOER acknowledges CONAHCYT for a graduate scholarship and financial support as an assistant to an SNII III level researcher}. GP is supported by the Portuguese Funda\c{c}\~{a}o para a Ci\^{e}ncia e Tecnologia (FCT) under Contracts UIDB/00777/2020, and UIDP/00777/2020, these projects are partially funded through POCTI (FEDER), COMPETE, QREN, and the EU. GP has a postdoctoral fellowship in the framework of UIDP/00777/2020 with reference BL154/2022\_IST\_ID. GZ would like to thank the MPP-Munich and DFG Exzellenzcluster 2181:STRUCTURES of Heidelberg University for support. 

\appendix

\section{Higgs sector rotations\label{ApendiceRotation}}

In the calculations carried out in this work, the following $R$ and $S$ rotation matrices were used:

\begin{equation}
	R=\left(\begin{matrix}0&0&{\widetilde{\alpha}}_4r_1&-{\widetilde{\alpha}}_3r_1\\{\widetilde{\alpha}}_1\frac{r_2}{r_1}&{\widetilde{\alpha}}_2\frac{r_2}{r_1}&-{\widetilde{\alpha}}_3\frac{r_1}{r_2}&-{\widetilde{\alpha}}_4\frac{r_1}{r_2}\\{\widetilde{\alpha}}_1&{\widetilde{\alpha}}_2&{\widetilde{\alpha}}_3&{\widetilde{\alpha}}_4\\{\widetilde{\alpha}}_2r_2&-{\widetilde{\alpha}}_1r_2&0&0\\\end{matrix}\right) \ \ \ \textrm{,} \ \ \
 	S=\left(\begin{matrix}0&{\widetilde{\beta}}_1\frac{s_2}{s_1}&{\widetilde{\beta}}_1&{\widetilde{\beta}}_2s_2\\0&{\widetilde{\beta}}_2\frac{s_2}{s_1}&{\widetilde{\beta}}_2&-{\widetilde{\beta}}_1s_2\\{\widetilde{\beta}}_4s_1&-{\widetilde{\beta}}_3\frac{s_1}{s_2}&{\widetilde{\beta}}_3&0\\-{\widetilde{\beta}}_3s_1&-{\widetilde{\beta}}_4\frac{s_1}{s_2}&{\widetilde{\beta}}_4&0\\\end{matrix}\right) ~,
  \label{RotationMatrices}
\end{equation}

\noindent where

\begin{equation}
	r_1=\left({\widetilde{\alpha}}_3^2+{\widetilde{\alpha}}_4^2\right)^{-\frac{1}{2}} \ \ \textrm{,} \ \ r_2=\left({\widetilde{\alpha}}_1^2+{\widetilde{\alpha}}_2^2\right)^{-\frac{1}{2}} \ \ \ \textrm{,} \ \ s_1=\left({\widetilde{\beta}}_3^2+{\widetilde{\beta}}_4^2\right)^{-\frac{1}{2}} \ \ \textrm{,} \ \ s_2=\left({\widetilde{\beta}}_1^2+{\widetilde{\beta}}_2^2\right)^{-\frac{1}{2}} ~.
\end{equation}

The parameters ${\widetilde{\alpha}}_i$ and ${\widetilde{\beta}}_i$ must satisfy the following relation to ensure that $R$ and $S$ are orthogonal matrices:

\begin{equation}
    \sum_{i=1}^{4}{\widetilde{\alpha}}_i^2=\sum_{i=1}^{4}{\widetilde{\beta}}_i^2=1 ~.
    \label{CondicionOrtogonalidad}
\end{equation}

The rotation matrices $R$ and $S$ (\ref{RotationMatrices}) are a particular case of a general rotation in four dimensions. However, in this work, it is useful to express three of the elements of row three of $R$ and three of the elements of column three of $S$ as independent parameters in order to associate the two Higgs sectors in a simple way.

\bibliographystyle{elsarticle-num}

\bibliography{All_Refs}

\end{document}